\documentclass[useAMS,usenatbib,usegraphicx]{mn2e}
\usepackage{multirow}
\usepackage{xcolor}
\usepackage{times}
\title[Polarization Alignments of Radio Quasars]{Polarization alignments of radio quasars in JVAS/CLASS surveys}
\author[V. Pelgrims and D. Hutsem{\'e}kers]{V. Pelgrims$^{1}$\thanks{E-mail:
pelgrims@astro.ulg.ac.be} and D.
Hutsem{\'e}kers$^{1,2}$\\
$^{1}$IFPA, AGO Dept., University of Li{\`e}ge, B4000 Li{\`e}ge, Belgium\\
$^{2}$AEOS, AGO Dept., University of Li{\`e}ge, B4000 Li{\`e}ge, Belgium}

\begin{document}

\date{Accepted 000. Received 000; in original form 2014 December 23}

\pagerange{\pageref{firstpage}--\pageref{lastpage}} \pubyear{2015}

\maketitle

\label{firstpage}

\begin{abstract}
We test the hypothesis that the polarization vectors of flat-spectrum radio sources (FSRS) in the JVAS/CLASS 8.4-GHz surveys are randomly oriented on the sky. The sample with robust polarization measurements is made of $4155$ objects and redshift information is known for $1531$ of them.
We performed two statistical analyses: one in two dimensions and the other in three dimensions when distance is available.
We find significant large-scale alignments of polarization vectors for samples containing only quasars (QSO) among the varieties of FSRS's.
While these correlations prove difficult to explain either by a physical effect or by biases in the dataset, the fact that the QSO's which have significantly aligned polarization vectors are found in regions of the sky where optical polarization alignments were previously found is striking.
\end{abstract}

%\nokeywords

\section{Introduction}
\label{sec:Intro}

Hutsem\'ekers et al. (\citeyear{Hutsemekers1998},~\citeyear{Hutsemekers-Lamy2001},~\citeyear{Hutsemekers-et-al2005}) have reported intriguing alignments of polarization vectors of quasars at optical wavelengths. The presence of these alignments may possibly be in tension with the well-accepted concordance model of cosmology as they involve correlations between sources separated by comoving distances at the Gpc-scale.

Paying particular attention to instrumental biases, \citet{Jackson-et-al2007} compiled the JVAS/CLASS 8.4-GHz sample of flat-spectrum radio sources (FSRS) with polarization position angle (PA) measurements. As they have shown, rotation measures of the polarization vectors induced by Faraday rotation at 8.4 GHz are too small to destroy information about the intrinsic PA's. Therefore, any observed correlation of PA's among sources can be thought to be intrinsic to the sources themselves.
Extracting from this sample 4290 FSRS's with polarized flux higher than 1 mJy,
\citet{Joshi-et-al2007} searched for systematic alignments of radio polarization vectors of the type reported at optical wavelengths by Hutsem\'ekers and collaborators, i.e.  at cosmological scales. Their analysis did not reveal such large-scale alignments at radio wavelengths.
From this claim, the recognized wavelength dependence of the polarization vector alignments has brought the model of axion-like particle (e.g. \citealt{Das-et-al2005},~\citealt*{Payez-Cudell-Hutsemekers2008}, \citealt*{Agarwal-Kamal-Jain2011}) as the favourite candidate to explain alignments at optical wavelengths. This model has however been observationally ruled out since it predicts non-negligible circular polarization which is not detected (\citealt{Hutsemekers-et-al2010} and \citealt*{Payez-Cudell-Hutsemekers2011}).

Beside this analysis, \citet{Tiwari-Jain2013} tested the uniformity of the polarization PA's 
considering roughly the same sample. They found significant evidences for alignments at distance scale of the order of 150 Mpc\footnote{Attention has to be paid regarding this scale as these authors defined the comoving distances assuming a redshift of one for all objects.}. As the correlations are found at different distance scales, their study does not contradict the analysis of \citet{Joshi-et-al2007}.
More recently, \citet{Shurtleff2014} studied the correlation of the PA's for sources grouped in circular regions of $24^{\circ}$ radius. While not very significant, he reported indication of PA's alignments in two regions of the sky.

Despite these analyses which involve different statistical tests and different samples which correspond to different cuts of the original dataset, the status of polarization PA correlation at radio wavelengths in not clear. Moreover, an analysis taking the redshift of the sources into account in still missing.
The redshift dependence being an important characteristic of the alignments of quasar polarization vectors at optical wavelengths, it seemed important to us to take it into account in the analysis of the radio sample. Especially if one seeks the same signature at radio wavelengths as at optical wavelengths.
Therefore, originally motivated by the redshift dependent analysis of this radio sample on the one hand, and by the recent and independent confirmation by \citet{Pelgrims-Cudell2014} of the polarization alignments at optical wavelengths, on the other hand, we devote this study to a careful analysis of the uniformity of the polarization PA's of FSRS's belonging to the JVAS/CLASS 8.4-GHz surveys.

We introduce the data samples which are studied throughout this work in Section~\ref{sec:DataSample}.
In Section~\ref{sec:VisibleWindows}, taking the redshift of the sources into account, we investigate the polarization PA distributions of the FSRS's located in regions of the sky where the optical polarization alignments are the most significant. Stimulated by the detection of alignment in one of these regions, we perform a complete analysis of the entire dataset in Section~\ref{sec:uniformity_in_JVAS}, with and without accounting for the redshift.
Having highlighted significant alignment signatures in the sample of quasars (hereafter, QSO), we search for their characterization in Section~\ref{sec:IdentifRegions}.
We finally summarize our results in Section~\ref{sec:DiscussionFinal}, bring arguments against and for the hypothesis of biases in the dataset and discuss a possible interpretation of the data.
We conclude in Section~\ref{sec:Conclusion} that the dataset of the polarization angle measurements of the JVAS/CLASS 8.4-GHz surveys are either not exploitable due to instrumental biases or, more excitingly, that we may have pinpointed large-scale alignments at radio wavelengths.
As we use little known statistical tests throughout this work, we found necessary to discuss them. Thereby, the descriptions are brought together in the Appendix in order to keep a fluent reading.

\section{Data sample}
\label{sec:DataSample}
The JVAS/CLASS 8.4-GHz catalogue is made of the JVAS (Jodrell-VLA Astrometric Survey) and the CLASS (Cosmic Lens All-Sky Survey) surveys that were gathered by \citet{Jackson-et-al2007} to build the largest catalogue of polarization measurements of compact radio sources, at that time, paying attention to biases on polarization measurements.
We refer to \citet{Jackson-et-al2007} and references therein for a complete description of the catalogue and the surveys.
In this catalogue, the total number of object having polarization measurements is $12.743$ (see the on-line catalogue\footnote{http://vizier.u-strasbg.fr/viz-bin/VizieR-3?-source=J/MNRAS/376/371}).
Adopting the prescription of \citet{Jackson-et-al2007} and \citet{Joshi-et-al2007}, we retain the sources for which the polarized flux is higher or equal to 1 mJy in order to keep significant polarization detections only and to obtain an unbiased sample. When there is more than one object in a radius of 1 arcsec on the sky, we select the object with the highest polarized flux. This selection, which also eliminates multiple measurements, leaves us with a sample size of $4265$ objects\footnote{If we only remove multiple measurements, we recover the source number of $4290$ studied by \citet{Joshi-et-al2007}. We nevertheless choose to add the above constraint for an efficient source separation.}.
Led by a selection criterion used at optical wavelengths \citep[see][for a discussion]{Hutsemekers1998}, we further constrain the sample asking that $\sigma_{\theta} \leq 14^{\circ}$, where $\sigma_{\theta}$ is the error on the polarization PA.
Out of the $4265$ sources, $4155$ satisfied the criterion. This sample, which we call $All$ in the reminder, constitutes the largest sample for which we have robust polarization PA measurements from the JVAS/CLASS 8.4-GHz surveys.

Using the NASA Extragalactic Database\footnote{http://ned.ipac.caltech.edu/} (NED), we identified a total of $3858$ sources. We first used the automated mode "Near-Object/Position List" with a search radius of $0.1$ arcsec. After manual clarification of multiple identifications, $3446$ objects were found.
For the $709$ objects left, we used a search radius of $0.5$ arcsec and found $412$ additional sources, after having again manually clarified the multiple identifications.
We stopped the procedure at this value of the search radius in order to ensure proper identifications.
Out of the $3858$ retrieved objects, $1531$ have spectroscopic, and thus reliable, measurements of redshift, $z$.

The use of NED also leads to the classification of the sources.
Table~\ref{tab:NED_classification} reports the number of sources identified for each class of FSRS as well as the number of these sources for which we have redshift information.
\begin{table}
\caption{The different samples.}
\label{tab:NED_classification}

\medskip

\begin{tabular}{llrc}
\hline
$z$					& Object Type			& $n$	& acronym\\
\hline
\multirow{5}{*}{no}	& All 				& 3858	& 	$-$		\\
					& QSOs 				& 1450 	& 	$QSO$	\\
					& Radio Sources 	& 1379 	& 	$RS$ 	\\
					& Galaxies 			& 381 	& 	$G$ 	\\
					& Other	Objects	 	& 648 	& 	$VO$ 	\\
\hline
\multirow{5}{*}{yes}& All				& 1531	&	$All(z)$\\
					& QSOs 				& 1325 	& 	$QSO(z)$\\
					& Radio Sources 	& 11 	& 	$-$		\\
					& Galaxies 			& 184 	& 	$-$		\\
					& Other	Objects 	& 11 	& 	$-$		\\
\hline
\end{tabular}

\medskip

Number ($n$) of the different source species for the objects retrieved in the NED database among the sample of $4155$ sources with reliable polarization PA measurements, with and without redshift information, $z$. The last column contains the acronyms used for the samples analysed in this work. The category named \textit{Other Objects} contains various species with small membership.
\end{table}
As it can be clearly seen, the QSO's represent $86 \%$ of the sample with redshift measurements. Hence, analyses and results involving samples with redshift information will mainly concern those objects.

\section{Uniformity of polarization PA's in JVAS/CLASS 8.4-GHz surveys}
\subsection{Regions of optical polarization alignments}
\label{sec:VisibleWindows}
In \citet{Hutsemekers1998}, \citet{Hutsemekers-et-al2005} and recently in \citet{Pelgrims-Cudell2014}, specific regions of the sky have been highlighted for which polarization PA's of quasars are found to be aligned at optical wavelengths. 
In the first studies, the two most significant regions were identified by eye and were called A1 and A3. In the latter and independent identification, \citet{Pelgrims-Cudell2014} used a more unbiased method and highlighted regions N2 and S2.
While less extended, the latter two regions were consistently found at similar locations in the 3-dimensional space.
Here, as the sky coverage of the radio surveys and the optical catalogue are different, we choose to consider the most extended regions (A1 and A3) to ensure an overlap as big as possible. Furthermore, those regions have been the subject of various studies in the past. They are delimited in right ascension, declination and redshift by:
\begin{itemize}
\item A1: $168^\circ \leq \alpha \leq 218^\circ$ ; $\delta \leq 50^\circ$ and $1.0 \leq z \leq 2.3$
\item A3: $320^\circ \leq \alpha \leq 360^\circ$ ; $\delta \leq 50^\circ$ and $0.5 \leq z \leq 1.5$
\end{itemize}
where $\alpha$ and $\delta$ refer to the right ascension and the declination of the sources, respectively.

\citet{Joshi-et-al2007} addressed the question of uniformity of the polarization PA's of FSRS's from the JVAS/CLASS 8.4-GHz surveys in these regions and reported no obvious alignment. However, they did not introduce the cuts in redshift and thus, only considered the \textit{windows} toward the A1 and A3 regions, defined by cuts in right ascension and declination only.
As the redshift of the sources is an important characteristic of the alignments of optical polarization PA's, we perform a new analysis of these regions.
It is nevertheless important to realize that the sky coverages of the JVAS/CLASS FSRS's and the studied quasars at optical wavelengths are different. In particular, the surveys at radio wavelengths do not contain data at $\delta<0^\circ$.

\medskip

To study the polarization PA distributions of the A1 and A3 regions, we use the Hawley--Peebles test (\citealt{Hawley-Peebles1975} and \citealt{Godlowski2012}) and the density test introduced in \citep{Pelgrims-Cudell2014}. We refer to Sections~\ref{HP_test} and~\ref{PCdensity_test} for a description of these statistical tests.
Considering samples of sources that are in a small region of the sky, these tests return the probability that the observed distribution of PA's is random and define the mean PA's ($\bar{\theta}$ and $\bar{\theta}_{\rmn{PC}}$, resp.) which are relevant only in the case of non-uniformity.

The Hawley--Peebles test analyses PA histograms. The number of bins is a free parameter. We decide to use $18$ bins of $10^\circ$ each, spanning the range $0^\circ \,-\,180^\circ$. This choice does not maximize the reported probabilities but is somehow justified by the fact that a bin width of $10^\circ$ corresponds approximately to twice of the mean of the errors of the polarization PA's under study.

The density test involves spherical caps of equal area. This area is fixed by the angular aperture of the cap ($2\eta$) which is the only parameter of the test. As discussed in the Appendix, it is useful to investigate a large range of values for $\eta$. Therefore, we explore here the range $2^\circ \,-\,90^\circ$ for $\eta$ and we report the probabilities $p_{min}$ and $p^\sigma$ corresponding to the value of $\eta$ for which $p^\sigma$ is the smallest.

Results of the tests applied to the sub-samples extracted from the $All(z)$ and the $QSO(z)$ samples are shown in Table~\ref{tab:HP-PC_A1A3}.
\begin{table*}
\begin{minipage}{105mm}
\caption{Test results in the A1 and A3 regions.}
\label{tab:HP-PC_A1A3}
\medskip
\begin{tabular}{@{}lcccc}
\hline
	& \multicolumn{2}{c}{A1 region} 	& \multicolumn{2}{c}{A3 region} 	\\
	& 						$All(z)$	& $QSO(z) $ & $All(z)$	& $QSO(z)$			\\
\hline
$n$ 						& $141$		& $139$		& $50$		& $45$				\\
$P_{\rmn{HP}}\,(\%)$				& $95.43$	& $93.15$	& $4.19$ 	& $0.96$			\\
$\bar{\theta}\,({}^\circ)$	& $-$		& $-$		& $64$		& $62$ 				\\
\hline
$p_{min}\,(\%)$ 					& $\geq4$	& $\geq5$	& $0.01$ & $7.2\,10^{-4}$	\\
$\eta\,({}^\circ)$ 			&$20\,-\,90$&$20\,-\,90$& $58$		& $56$				\\
$p^{\sigma}\,(\%)$ [$n_{sim}$]	& $\geq13$ [$10^2$] 	& $\geq10$ [$10^2$]
							& $0.07$ [$5\,10^4$] 	& $4.0\,10^{-3}$ [$5\,10^4$]\\
$\bar{\theta}_{\rmn{PC}}\,({}^\circ)$	& $-$	& $-$		& $68$ 	& $67$		 	\\
\hline
\end{tabular}

\medskip

Results of the Hawley--Peebles and the density tests performed on the sub-samples corresponding to the A1 and A3 regions of optical polarization alignments. Sub-samples are obtained from both the $All(z)$ and the $QSO(z)$ samples. $n$ is the size of the sub-samples, $P_{\rmn{HP}}$ is the probability given by the Hawley--Peebles test that the PA's are drawn from a uniform parent distribution and $\bar{\theta}$ is the mean polarization PA returned by this method. $p_{min}$ is the local probability obtained with the density test for the half-aperture angle $\eta$ to which corresponds the minimum global probability $p^{\sigma}$ computed with $n_{sim}$ random simulations. The mean angle $\bar{\theta}_{PC}$ is computed as explained in the text. Probabilities are given in percent.
\end{minipage}
\end{table*}
The hypothesis of uniformity of the polarization orientations is rejected at the level of at least $95\%$ in the A3 region but not in the A1 region\footnote{At first glance, the fact that we found alignment in the A3 region but not in the A1 region could be caused by a bad spatial overlap between radio and optical data in the A1 region. But the overlapping is not better in the A3 region.}.
The discovery of this alignment is very intriguing, especially given the claim of \citet{Joshi-et-al2007}.
To reconcile the analyses, we extracted the A3 window from the sample of $4155$ FSRS's. Applying our tests to this sub-sample of $385$ sources, we did not find any evidence for alignment and thus confirm the negative result of \citet{Joshi-et-al2007} (see Table~\ref{tab:A3window} and Section~\ref{sec:PrelResults} for further related discussions).

\medskip

As a result, we contradict the claim of \citet{Joshi-et-al2007} which states that no alignment is present at radio wavelengths (8.4 GHz) inside the regions where the optical polarization vectors are found to be coherently oriented.
This contradiction is likely due to the redshift cuts.

\medskip

It is of interest to find out whether the alignment tendency observed in this region is an isolated structure inside the sample of FSRS's or if it is part of a major trend which was not recognized by earlier works.
To this end, we shall address the question of uniformity of the polarization PA's for the complete JVAS/CLASS 8.4-GHz surveys, without restriction on the sky location, taking redshift into account and considering the subdivision of the sample into the different species.

\subsection{Full sky coverage}
\label{sec:uniformity_in_JVAS}
To study the uniformity of polarization angle distributions for samples of sparse and non-uniformly scattered sources on the celestial sphere, tools have been developed by \citet{Hutsemekers1998}, \citet*{Jain-Narain-Sarala2004}, \citet{Hutsemekers-et-al2005}, \citet{Joshi-et-al2007} and by \citet{Pelgrims-Cudell2014}.
The so-called S and Z tests, established by \citet{Hutsemekers1998} and modified by \citet{Jain-Narain-Sarala2004} to obtain coordinate-invariant statistics, are appropriate to assess the probability that the distributions of polarization PA's of local groups are due to statistical fluctuations considering the overall sample. We use these tests in this section.
The intrinsically coordinate-invariant statistical test introduced in \citet{Pelgrims-Cudell2014} being more useful for the characterization of correlations is used in Section~\ref{sec:IdentifRegions}.
We do not use the other statistical tests as they are coordinate-dependent, this dependence growing with the angular distance between sources. They are thus not adequate to test the uniformity of the PA distribution over large scales.

The coordinate-invariant S and Z tests are extensively discussed in Section~\ref{SandZ_tests}.
These nearest-neighbour tests compute the probability that the polarization PA's
are uniformly distributed in spatially defined groups of objects, making use of Monte Carlo simulations.
For each realization, the PA's are reshuffled among the sources of the entire sample and a statistics is computed for each group of $n_v$ nearest neighbours. The percentage of Monte Carlo simulations having an average statistic ($S_D$ or $Z_c$) as extreme as the one of the data defines the significance level (S.L.) of the test, i.e., the probability that the observed PA correlations inside groups can be attributed to statistical fluctuations in the entire sample.
To evaluate the S.L., we set the number of random simulations to $1000$, except contraindication.
For the samples of Table~\ref{tab:NED_classification}, we explore  a wide range of values of the parameter $n_v$ (see Section~\ref{nv-param} for the motivation of doing this). We span the range $4 \,-\,400$ (with steps of $2$ and $20$ for ranges $4\,-\,18$ and $20\,-\,400$, respectively), except for the sample of galaxies where we stop at $n_v=200$ for obvious reasons.

\medskip

We first consider samples for which reliable redshift measurements are available. For these samples, both 2- and 3-dimensional analyses are applied, i.e. defining nearest-neighbour groups on the celestial sphere or in the 3D comoving space respectively. We then turn to the 2-dimensional analysis of the samples of Table~\ref{tab:NED_classification} which are not constrained by redshift.
For convenience, we give in Table~\ref{tab:Summary_SZtest} a summary of the results of these two tests applied to all the samples.

\subsubsection{Samples with redshift measurements}
\label{SZ_3DAnalysis}
\begin{figure}
\begin{center}
\includegraphics[width=\linewidth]{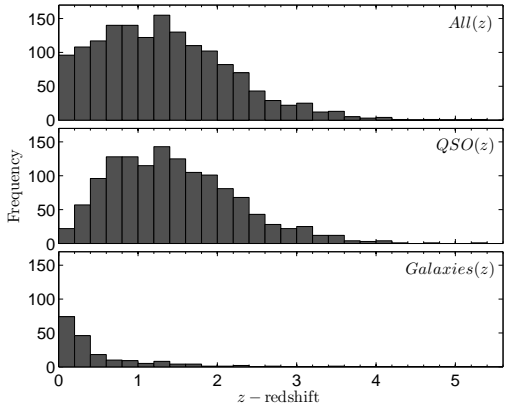}
\caption{Redshift distributions of the sample $All(z)$ and its sub-samples of QSO's and galaxies.}
\label{fig:zDistrib}
\end{center}
\end{figure}

As already mentioned, the sample of $1531$ sources for which reliable redshift measurement are available is composed at $86\%$ by QSO's. The second important population of this sample is that of galaxies.
The redshift distributions of these samples are shown in Fig.~\ref{fig:zDistrib}. Of course, the redshift distributions of the sample $QSO(z)$ and the one of galaxies do not follow the same trend.

\begin{figure}
\begin{center}
\begin{tabular}{@{}c}
\includegraphics[width=\linewidth]{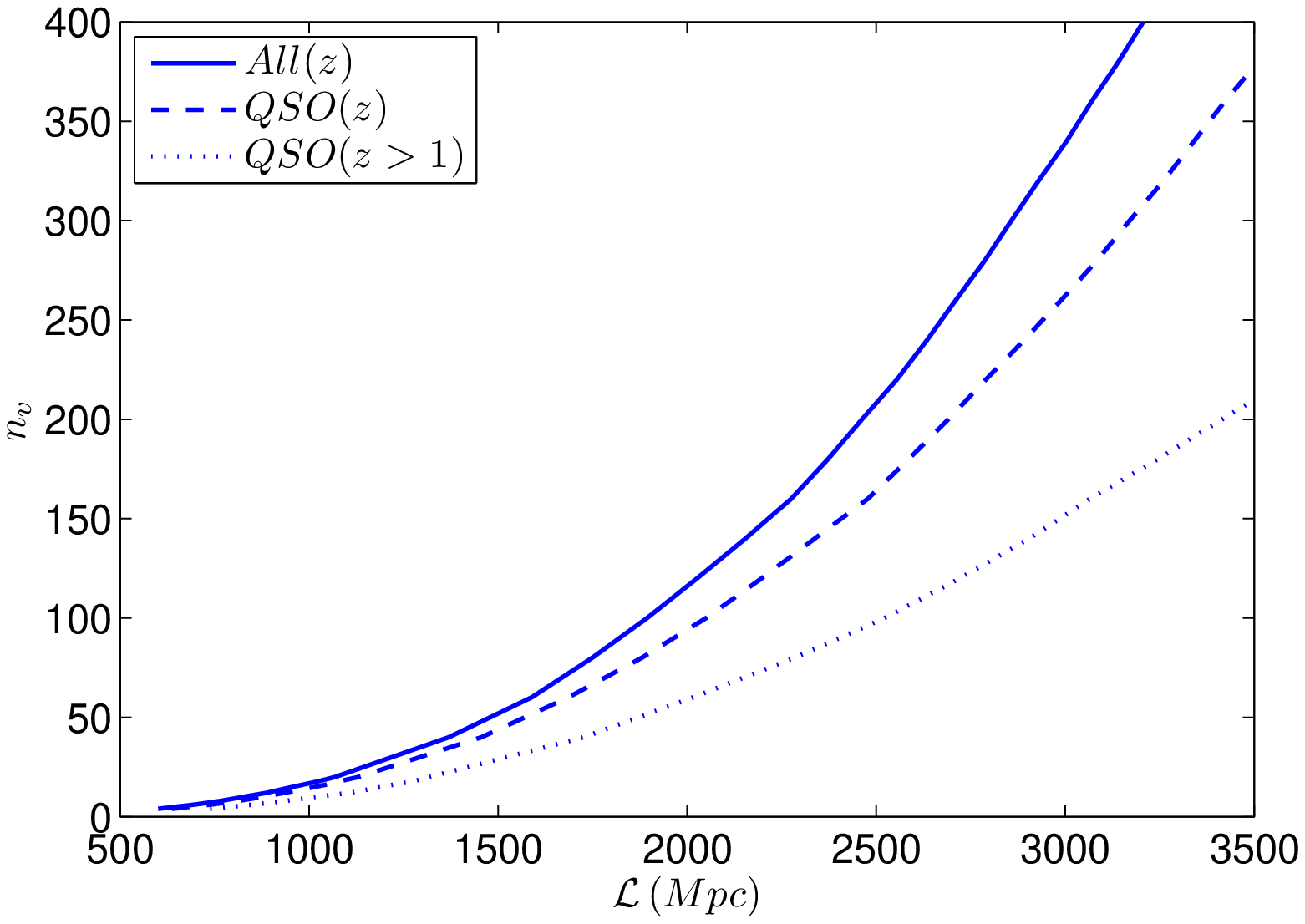}		\\
\includegraphics[width=\linewidth]{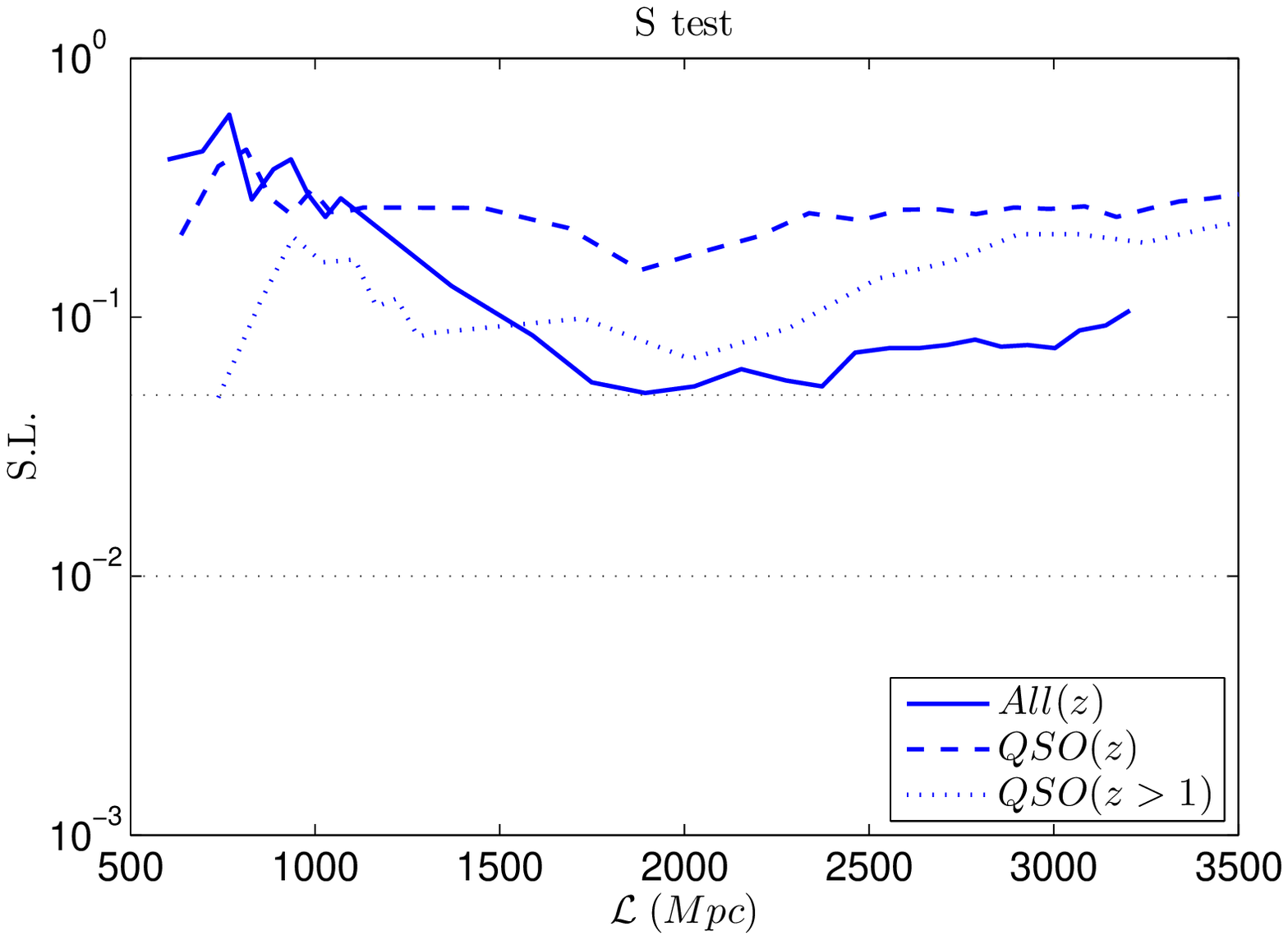}	\\
\includegraphics[width=\linewidth]{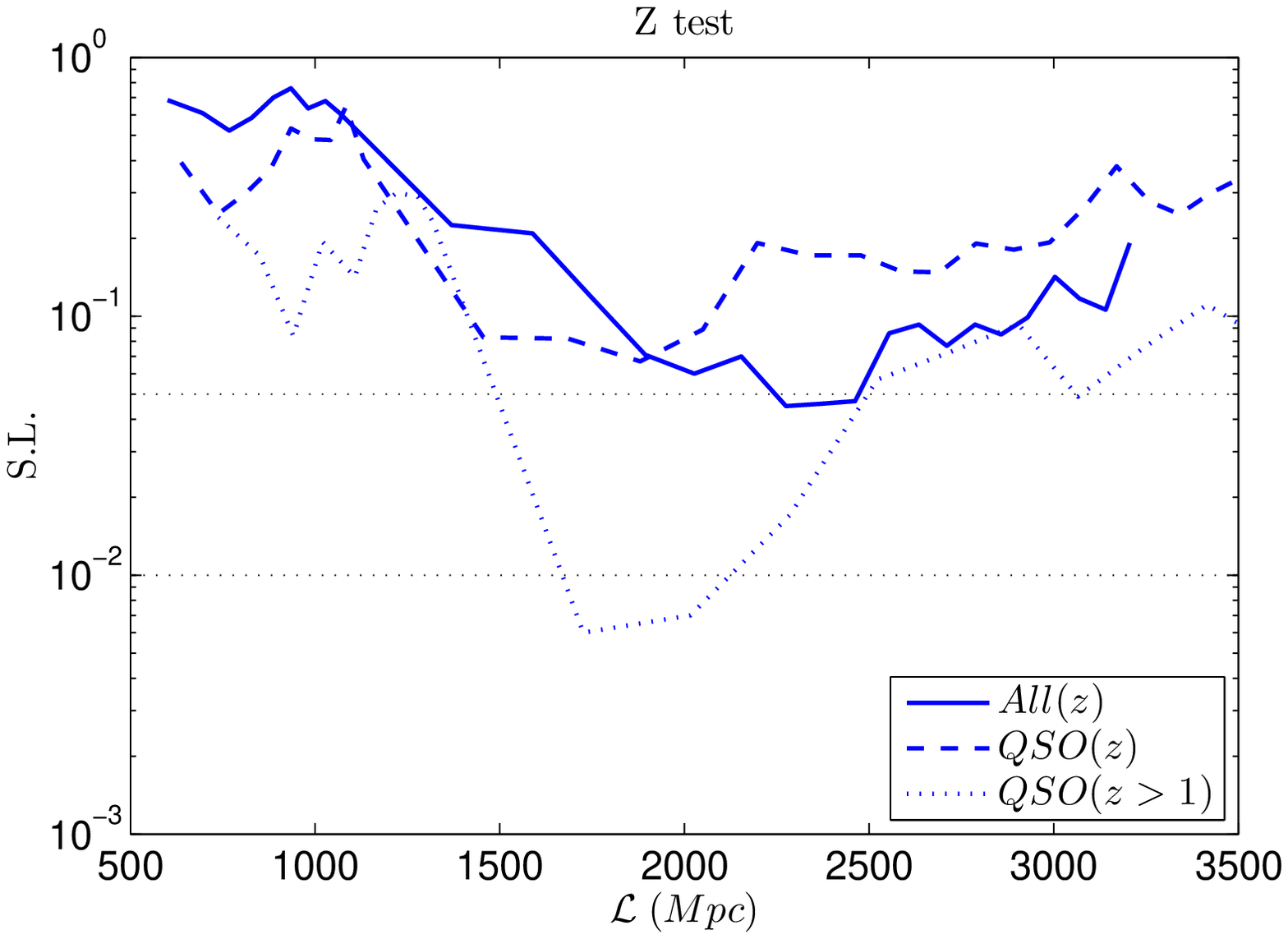}	\\
\end{tabular}

\caption{3D analysis of the samples with redshift measurements. Top: Relation between the parameter $n_v$ and the typical comoving separation $\mathcal{L}$ in Mpc for the samples $All(z)$, $QSO(z)$ and $QSO(z>1)$.
Middle and Bottom: Significance level obtained with the S and Z tests (resp.) as a function of the typical comoving distance $\mathcal{L}$ for the three samples. The $5\%$ and $1\%$ S.L. are indicated.}
\label{fig:3Danalysis-zSample}
\end{center}
\end{figure}

\paragraph{3-dimensional analysis.}
We discuss our results in term of a typical comoving distance $\mathcal{L}$ instead of the free parameter $n_v$ of the statistical tests. This typical scale is defined as being the median of the comoving distances between each central object and its $n_v$-\textit{th} nearest neighbour, the median being evaluated over the full sample under consideration. The line-of-sight comoving distances are computed assuming a flat Universe with the cosmological parameters: $\Omega_{M}=0.31$ and $H_0 = 68 \, \rmn{ km \, s}^{-1} \rmn{Mpc}^{-1}$.
We show in Fig.~\ref{fig:3Danalysis-zSample} (Top) the relation between the parameter value $n_v$ and the typical comoving distance for the samples we analyse in 3 dimensions.
We applied the statistical tests to the sample of $1531$ objects and to the subcategory of QSO, namely $All(z)$ and $QSO(z)$. We also considered the high redshift part of the latter, imposing $z>1$. This restricted sub-sample is populated by $894$ sources and is denoted $QSO(z>1)$.

\medskip

Results of the S and Z tests are shown in Fig.~\ref{fig:3Danalysis-zSample} (Middle and Bottom, respectively).
We did not find any significant evidence ($\rmn{S.L.} < 5 \%$) over a wide range of value of $n_v$ (or $\mathcal{L}$ ) for alignment of the polarization PA's in the sample $All(z)$ and $QSO(z)$.
However a redshift dependence is possibly detected with the Z test as suggested in Fig.~\ref{fig:3Danalysis-zSample} (Bottom).
Indeed, in the high redshift QSO sample, correlations of polarization PA's of sources inside groups of typical comoving radius of about  $2$ Gpc show a probability less than $1\%$ of being due to statistical fluctuations.

\begin{figure}
\begin{center}
\begin{tabular}{@{}c}
\includegraphics[width=\linewidth]{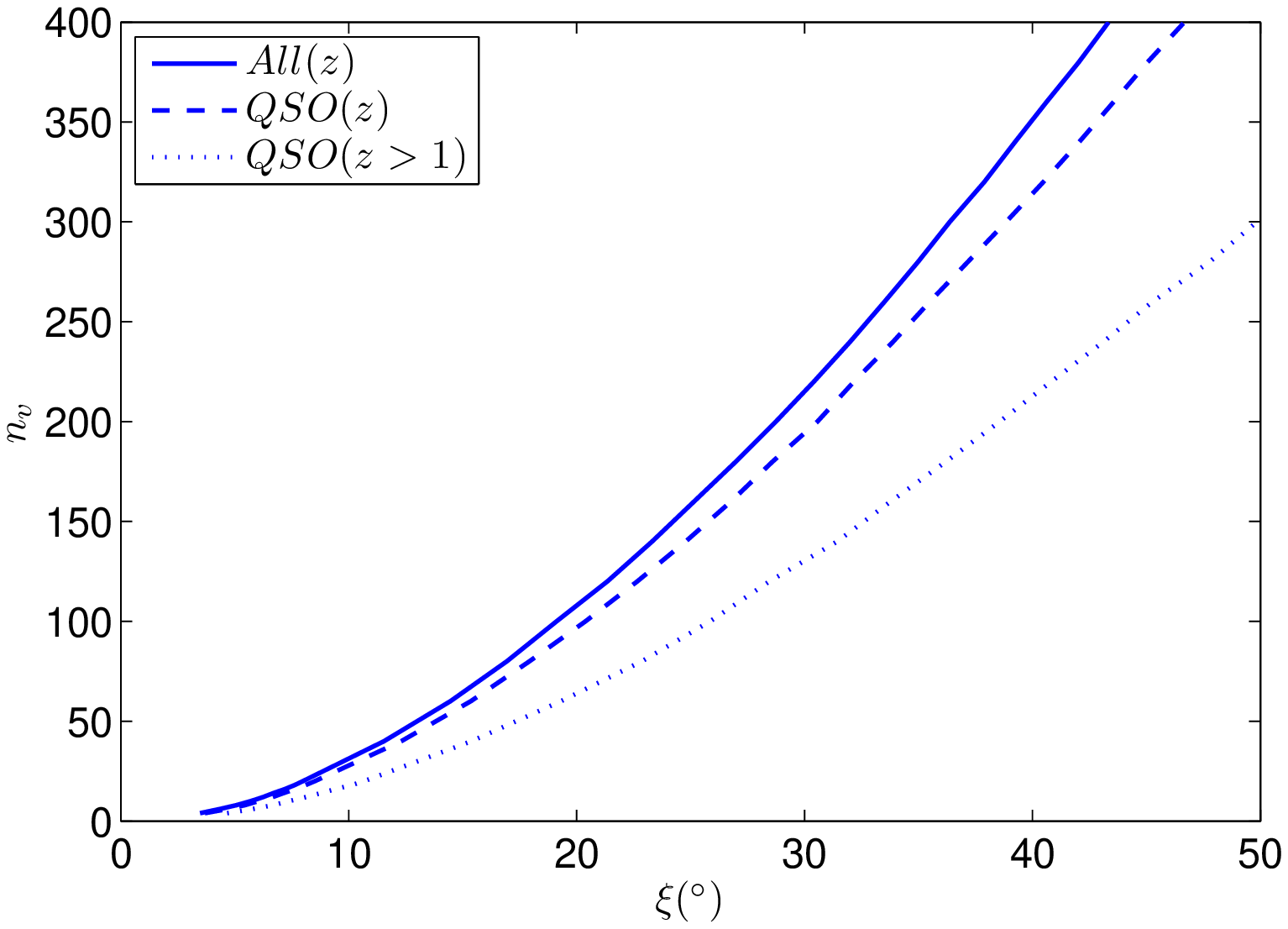}	\\
\includegraphics[width=\linewidth]{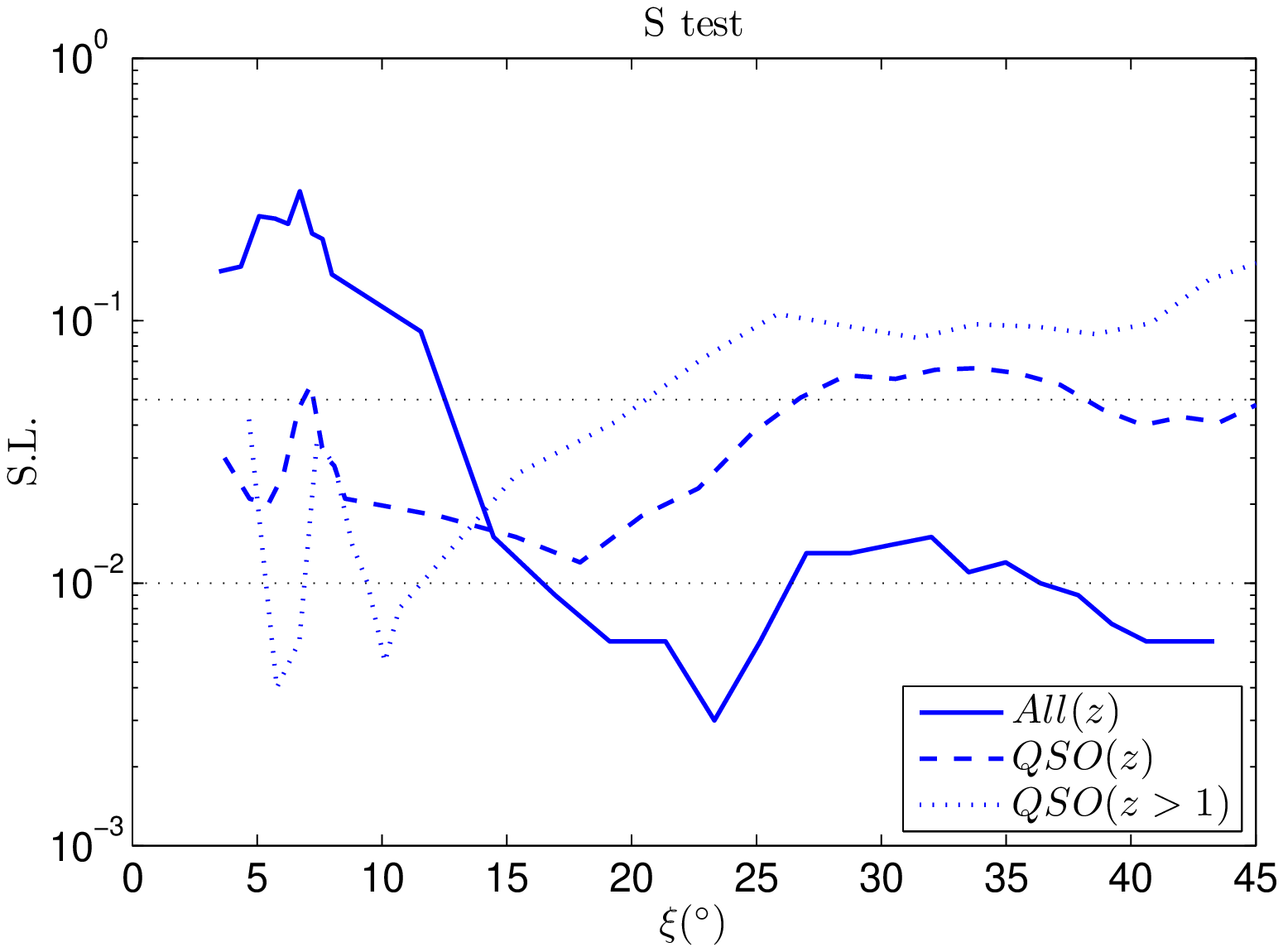}	\\
\includegraphics[width=\linewidth]{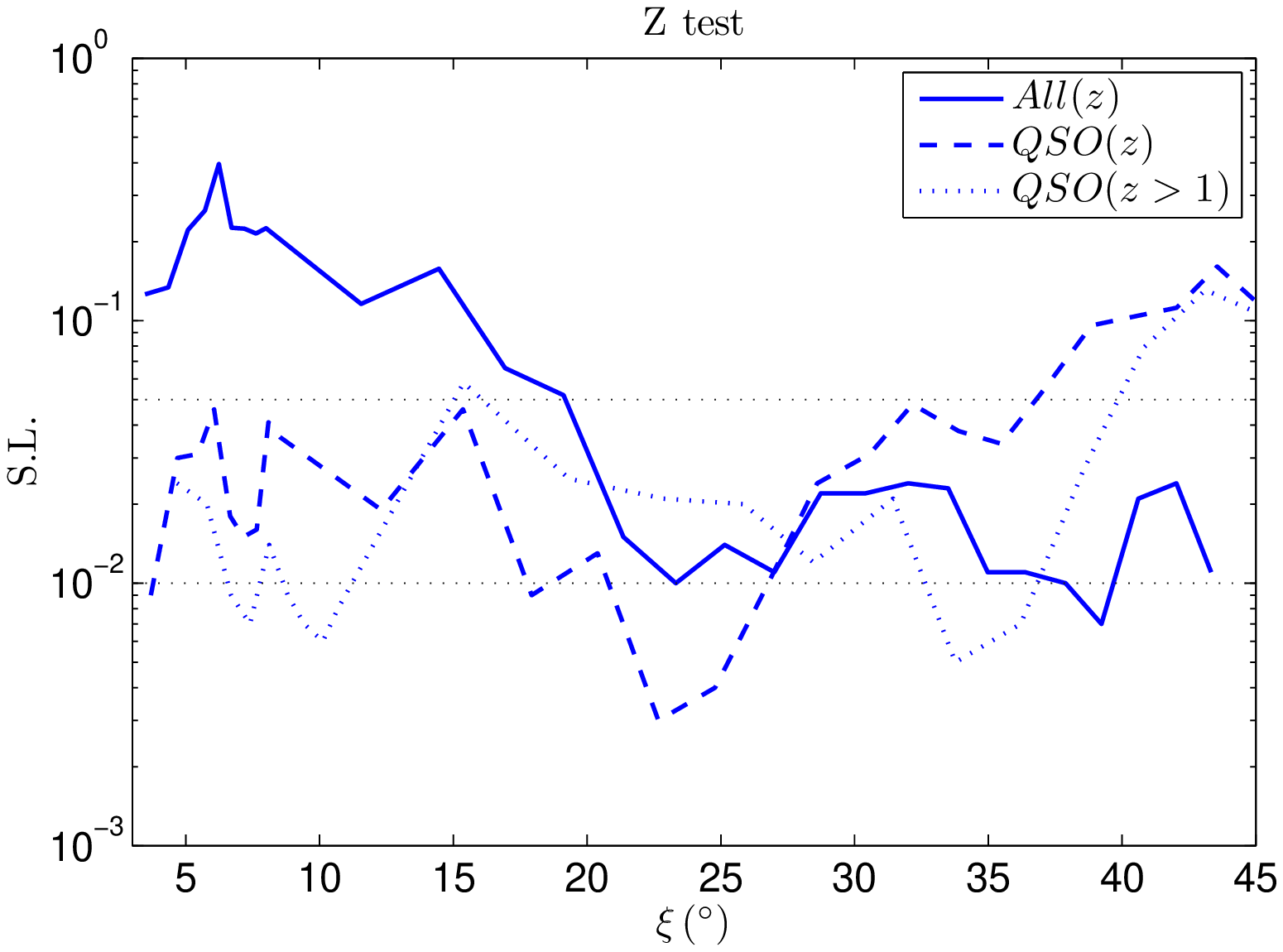}	\\
\end{tabular}

\caption{2D analysis of the samples with redshift measurements. Top: Relation between the parameter $n_v$ and the typical angular separation $\xi$ in degree for the samples $All(z)$, $QSO(z)$ and $QSO(z>1)$.
Middle and Bottom: Significance level obtained with the S and Z tests (resp.) as a function of the typical angular separation $\xi$ for the three samples. The $5\%$ and $1\%$ S.L. are indicated.}
\label{fig:2Danalysis-zSample}
\end{center}
\end{figure}

\paragraph{2-dimensional analysis.}
For the 2-dimensional analysis, the radial coordinate of the sources is fixed to $r=1$, even though redshift measurements are available. We discuss our results in term of the typical angular separation $\xi$. The latter is defined as being the median of the angular separation between each object of the sample and its $n_v$-\textit{th} nearest neighbour. Fig.~\ref{fig:2Danalysis-zSample} (Top) shows the relation between $\xi$ and $n_v$ for the three samples.

We show the dependence of the S.L. on the typical angular separation $\xi$ in Fig.~\ref{fig:2Danalysis-zSample}.
Significant correlations ($\rmn{S.L.}< 5\%$ over a wide range of $\xi$ value) of the polarization PA's inside groups is observed for the three samples ($All(z)$, $QSO(z)$ and $QSO(z>1)$) although the minima occur at different typical angular separations.
Indeed, for the S test, the sample $All(z)$ shows its minimum S.L. at $0.3\,\%$ for $ \xi \approx 23^\circ $ and $QSO(z)$ shows a small dip for the range of $\xi \approx 8\,-\,26 ^\circ$ with a minimum $\rmn{S.L.}=1.2\,\%$ for $\xi = 18 ^\circ$. The high redshift part of the QSO sub-sample ($QSO(z>1)$) exhibits values of the S.L. below $1\%$ for smaller angular separation ($\xi \la 10^\circ$). These features are somehow confirmed by the Z test as seen from Fig.~\ref{fig:2Danalysis-zSample} (Bottom). For this test, the minimum S.L. value of the sample $QSO(z)$ is found to be as low as $0.3\,\%$ for $\xi \approx 23^\circ$ and the sample $QSO(z>1)$ shows S.L. below $1\%$ for $\xi\leq10^\circ$ with an additional dip around $\xi=34^\circ$.

\subsubsection{Full samples with different object types}
\label{SZ_2DAnalysis}
\begin{figure}
\begin{center}
\begin{tabular}{@{}c}
\includegraphics[width=\linewidth]{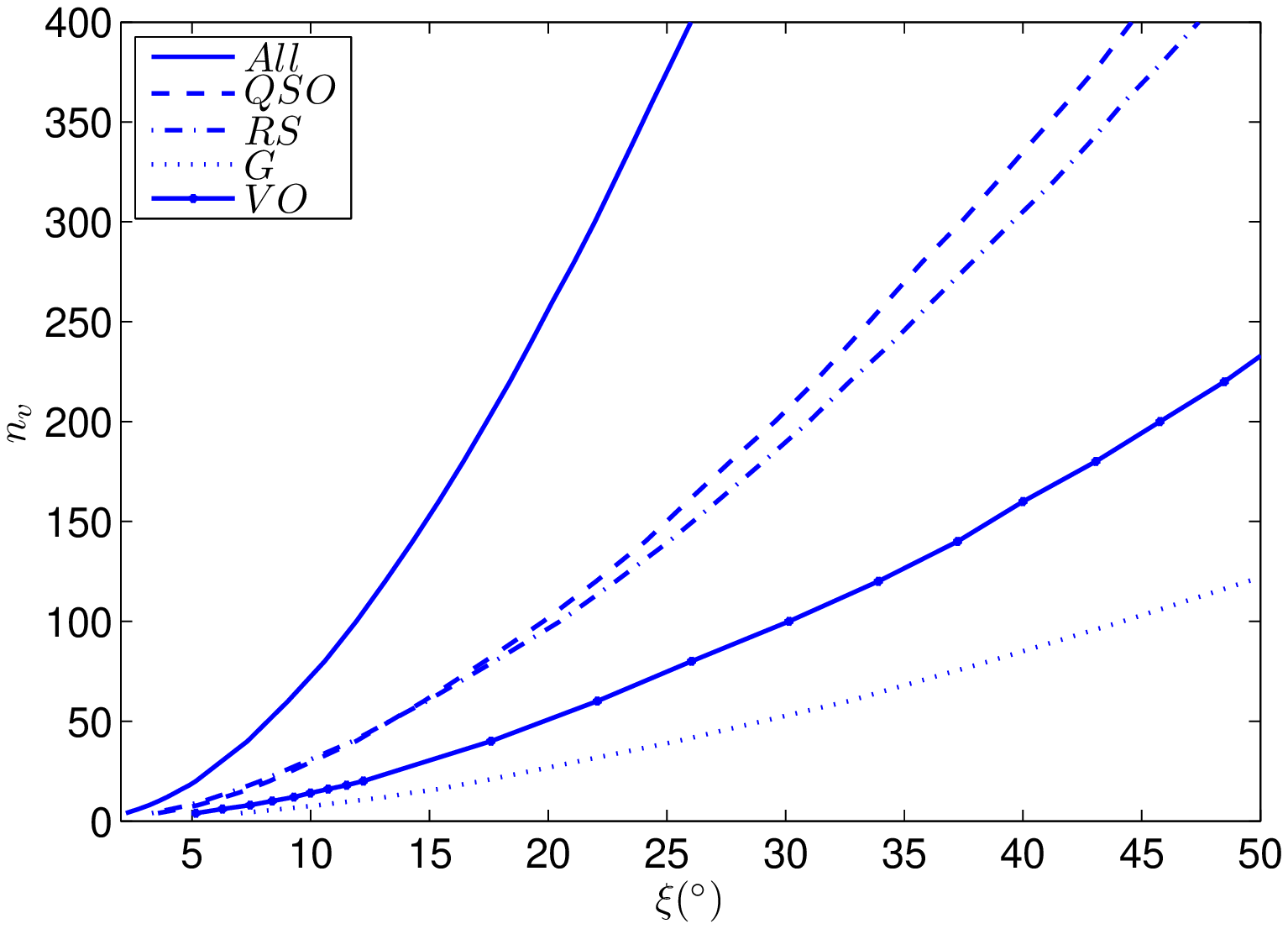}	\\
\includegraphics[width=\linewidth]{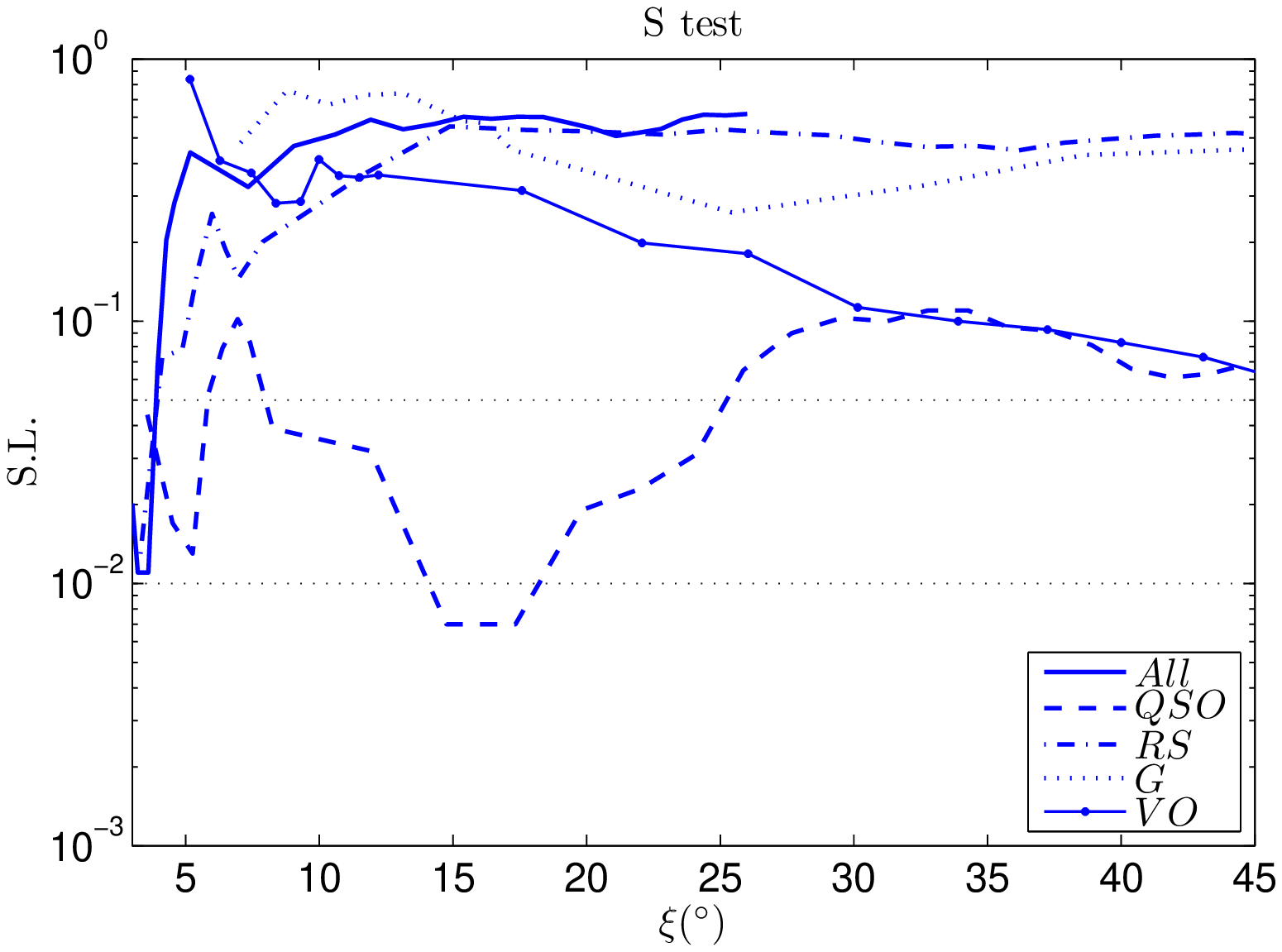}	\\
\includegraphics[width=\linewidth]{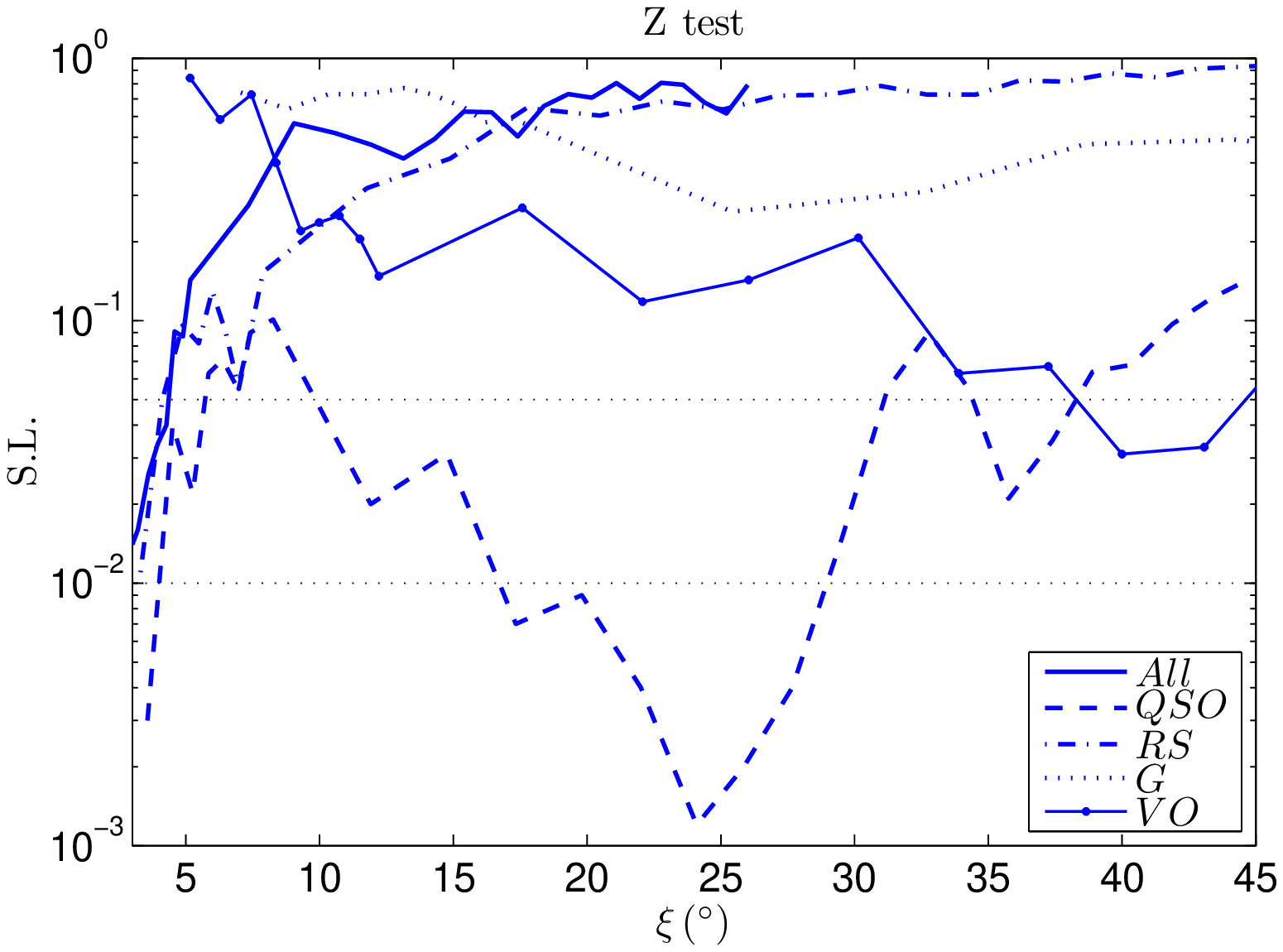}	\\
\end{tabular}

\caption{2D analysis of the full samples. Top: Relation between the parameter $n_v$ and the typical angular separation $\xi$ for the samples $All$, $QSO$, $RS$, $G$ and $VO$ of Table~\ref{tab:NED_classification}.
Middle and Bottom: Significance level obtained with the S and Z tests (resp.) as a function of the typical angular separation $\xi$ for the five samples. Note that the S.L. value of the sample $QSO$ for $\xi \approx 24^\circ$ ($n_v=140$) has been computed with $10^{4}$ random simulations.}
\label{fig:2Danalysis-fullSample}
\end{center}
\end{figure}

Analysing samples with redshift measurements in two dimensions, we have found significant correlations ($\rmn{S.L.}<5\%$ with minima $<1\%$). It is therefore interesting to also perform the 2-dimensional analysis on the other samples of Table~\ref{tab:NED_classification}, i.e. on the samples not restricted by the availability of the redshift of the sources.
We thus consider the sample of $4155$ sources as well as its four sub-samples with different object types.
We show in Fig.~\ref{fig:2Danalysis-fullSample} (Top) the relations between $n_v$ and $\xi$ for these five samples.
The results of the S and Z tests are shown in Fig.~\ref{fig:2Danalysis-fullSample} (Middle and Bottom respectively).

For small values of $n_v$ (from $6$ to $10$), we found indications of alignments in the sample $All$ as the S and Z tests return S.L. values at the percent level ($1.1 \%$ and $1.2 \%$).
These indications of alignments are reminiscent of the correlations highlighted by \citet{Tiwari-Jain2013} at the scale of $\sim 150\,\rmn{Mpc}$. The reasons why we found less convincing correlations are likely that (\textit{i}) we consider a different
sample\footnote{When we built the dataset from the JVAS/CLASS catalogue, we removed duplicate measurements while \citet{Tiwari-Jain2013} did not (Jain (private communication, 2015)).} and (\textit{ii}) that we use a different statistics.

For large values of $n_v$, alignments are detected with S.L. below $5\%$ over a wide range of $\xi$ only for the sample $QSO$.
The S.L. of the S test applied to $QSO$ exhibits a dip for $n_v$ between $40$ and $140$, reaching the value of $0.7\,\%$ for $n_v=60$ and $80$. The range of typical angular separations involved in these correlations is $\xi\approx12^\circ\,-\,24^\circ$, with stronger correlations for the range $14.5^\circ\,-\,17.5^\circ$.
The Z test exhibits a large dip for the range $n_v = 40\,-\,200$ with the minimum S.L. of $ 0.12\,\%$ for $n_v=140$, implying correlations at $\xi \approx 24^\circ$.
Those correlations of polarization PA's involve QSO's separated by large distances on the celestial sphere and confirm the detection made in the sample $QSO(z)$ (with redshift measurements) in Section~\ref{SZ_3DAnalysis}.

For such angular scales, the distributions of the polarization PA's of the other samples ($All$, $RS$, $G$ and $VO$) are  in good agreement with the hypothesis of uniformity.
Let us emphasize that these large-scale correlations are not observed for the category of radio sources ($RS$), even though the sample size is comparable to the one of $QSO$ (see Table~\ref{tab:NED_classification}).

\subsection{Preliminary results}
\label{sec:PrelResults}
So far, studying the polarization PA's of different samples drawn from the JVAS/CLASS 8.4-GHz surveys, we have found significant alignments in some of these samples; first, in one of the regions where the optical polarization vectors were found to be aligned and second, in the QSO all-sky survey.

\medskip

Regarding the A3 region, a few reasons might lead to the differences between our findings and the conclusions of \cite{Joshi-et-al2007}.
As already mentioned, when these authors analysed the so-called A1 and A3 regions, they did not constrain their sample with regard to the redshift, which is an important characteristic of the optical polarization alignments. They applied their \textit{Nearest Neighbour Test} to the full sample restricted to the sky \textit{window} of the A3 region, i.e. introducing cuts in right ascension and declination only \citep[see Section~3.3 and~5 of][]{Joshi-et-al2007}.
The first possible cause of divergent results is a redshift dependence of the alignments at radio wavelengths as at optical wavelengths, so that taking all the sources in the window regardless of their redshift blurs the alignment.
Alternatively, the comparison of the last two columns of Table~\ref{tab:HP-PC_A1A3} suggests that alignments are more pronounced for QSO compared to the other types of objects.

In order to test these two scenarios, we performed an analysis of the samples obtained by imposing the A3 window cut on the different samples of Table~\ref{tab:NED_classification}.
Results of the Hawley--Peebles test and the density test of Pelgrims and Cudell are shown in Table~\ref{tab:A3window}.
\begin{table*}
\begin{minipage}{90mm}
\caption{Test results with the A3 window cut on $QSO(z)$, $All(z)$, $QSO$ and $All$ samples.}
\label{tab:A3window}
\medskip
\begin{tabular}{@{}lcccc}
\hline
							&		\multicolumn{3}{c}{A3 window cut on}		\\
							& $QSO(z)$				& $All(z)$	
							& $QSO$					& $All$		\\
\hline
$n$ 						& $100$					& $115$	
							& $114$					& $385$		\\
%$P_{\rmn{HP}}_{sim}$		& $0.38\%$				& $1.16\% (5 10^3)$	
%							& $1.15\%$				& $29.8\%$	\\
$P_{\rmn{HP}}\,(\%)$				& $0.36$			 	& $1.27$ 
							& $1.19$				& $29.6$	\\
$\bar{\theta}\,({}^\circ)$	& $68$ 					& $72$ 		
							& $68$ 					& $-$		\\
\hline
$p_{min}\,(\%)$ 					& $7.0\,10^{-4}$		& $1.9\,10^{-3}$ 	
							& $5.9\,10^{-3}$		& $0.14$	\\
$\eta\,({}^\circ)$ 			& $52$					& $52$
					 		& $52$		 			& $68$		\\
$p^{\sigma}\,(\%)$ [$n_{sim}$]	& $0.02$ [$5\,10^4$]	& $0.01$ [$10^4$]
							& $5.4\,10^{-2}$ [$5\,10^4$]	& $1.5$ [$10^3$]	\\
$\bar{\theta}_{\rmn{PC}}\,({}^\circ)$	& $69$		& $70$
							& $68$					& $59$		\\
\hline
\end{tabular}

\medskip

Same as Table~\ref{tab:HP-PC_A1A3} but for the sub-samples obtained by application of the A3 window on the samples of $1325$ QSO's with redshift, $1531$ sources with redshift, $1450$ QSO's regardless of the redshift information, and $4155$ flat-spectrum radio sources.
\end{minipage}
\end{table*}
Correlations between polarization PA's are observed when we consider the A3 window cut of the samples $QSO$, $QSO(z)$ and $All(z)$, but no deviation from uniformity is detected for the A3 window cut of $All$, in agreement with the result of \citet{Joshi-et-al2007}.
Comparison of the last two columns of Table~\ref{tab:A3window} teaches us that adding the other species to the $QSO$ sample completely blurs the alignments.
This simple observation argues for the scenario in which the species selection is at the origin of the detection of the correlations. This scenario is reinforced when we consider the A3 window cut of the sample $RS$. For this sub-sample of $138$ objects, we found that the distribution of the polarization PA's is in agreement with the hypothesis of uniformity.
Comparison of Tables~\ref{tab:HP-PC_A1A3} and \ref{tab:A3window} does not allow us to conclude on a possible redshift dependence of the polarization alignments.
This is partially due to the lack of redshift information for species other than QSO's.

\medskip

In the full sample we highlighted alignments involving sources separated by typical angular scales of about $20^\circ$.
For the samples with redshift measurements, the large-scale correlations are observed to be more significant with the 2-dimensional analysis than with the 3-dimensional one.
Considering samples that are not limited by the redshift availability, we also pinpointed that the large-scale correlations mainly concern the category of QSO as it was already suggested during the study of the A3 window in Section~\ref{sec:VisibleWindows}.

\medskip

As a conclusion, we find that the polarization PA's of the JVAS/CLASS 8.4-GHz surveys show correlations in groups of QSO's with an angular radius of about $20^\circ$. The significance level at which these correlations can be attributed to statistical fluctuations in the sample of QSO is found to be as low as $\sim 0.1\%$ for $\xi \approx 24^\circ$ with the Z test.
\begin{table*}
\begin{minipage}{120mm}
\caption{Summary of S and Z test results.}
\label{tab:Summary_SZtest}
\begin{tabular}{@{}l|cccc|ccc}
\hline
	& \multicolumn{3}{|c|}{S} 	& \multicolumn{3}{c}{Z} 	\\
\hline
	3D		&	min(S.L.) ($\%$)&	$\mathcal{L}$ (Gpc)	&	$n_v$	&	 &
				min(S.L.) ($\%$)&	$\mathcal{L}$ (Gpc)	&	$n_v$		\\
\hline
$All(z)$	&	$-$			&	$-$				&	$-$		&	& 
					$-$			&	$-$					&	$-$			\\
$QSO(z) $ 	&	$-$			&	$-$				&	$-$		&	& 
					$-$			&	$-$					&	$-$			\\
$QSO(z>1)$	&	$-$			&	$-$				&	$-$		&	& 
					$0.6$		&	$\sim 1.7$			&	$40$		\\
\hline
	2D		&	min(S.L.) ($\%$)&$\xi \,({}^\circ)$	&	$n_v$	&	 &
				min(S.L.) ($\%$)&	$\xi \,({}^\circ)$	&	$n_v$		\\
\hline
$All(z)$	&	$0.3$	&	$\sim 23$		&	$140$	&	& 
					$1.0$		&	$\sim 23$			&	$140$		\\
$QSO(z) $ 	&	$1.2$	&	$\sim 18$		&	$80$	&	& 
					$0.3$	&	$\sim 23$			&	$120$		\\
$QSO(z>1)$	&	$0.5$	&	$\sim 10$		&	$18$	&	& 
			$0.6$ / $0.5$&	$10$ / $34$	&	$18$ / $160$		\\
\hline
$All$		&	$1.1$	&	$3\,-\,4$		&	$8\,-\,10$	&	& 
					$1.2$	&	$\sim 3$			&	$6\,-\,8$	\\
$QSO$		&	$0.7$	&	$14.5\,-\,17.5$	&	$60\,-\,80$	&	& 
					$0.12\,{}^*$	&	$\sim 24$			&	$140$		\\
$RS$		&	$1.3$	&	$\sim 3$		&	$4$			&	& 
					$1.1$	&	$\sim 3$			&	$4$			\\
$G$			&	$-$			&	$-$				&	$-$			&	 &
					$-$			&	$-$					&	$-$			\\
$VO$		&	$-$			&	$-$				&	$-$			&	 &
					$3.0$ 	&	$\sim 40$			&	$160$		\\
\hline
\end{tabular}

\medskip

Summary of the application of the S and Z statistical tests to all samples of Table~\ref{tab:NED_classification}. For each test, we report the value of the minimum S.L. with the corresponding $n_v$ parameter and its attached typical scale ($\mathcal{L}$ or $\xi$ for the 3- or 2-dimensional analysis, resp.). We only show results when the S.L. of the sample is found to be below the threshold of $5\%$ for a wide range of $n_v$. All S.L. have been evaluated with $1000$ Monte Carlo simulations except the smallest one (marked by an asterisk) for which we had to use $10000$ simulations.
\end{minipage}
\end{table*}

\section{Identification of regions of aligned polarizations}
\label{sec:IdentifRegions}
For the correlations highlighted in the previous section, it would be of interest to figure out if the alignments detected at typical scales of $\xi \approx 15^\circ\,-\,25^\circ$ are due to a global trend across the whole sky coverage of the survey or if they are prominent in some regions of the sky, as it seems to be the case at optical wavelengths.

To this end, we proceed to the identification of the groups of sources with distributions of polarization PA's that show significant departure from uniformity. The fact that these groups are clustered in space or not tells us whether the correlations of polarization orientations are due to well localized objects or to a general trend.
This identification can, inter alia, be achieved with the help of the S and Z tests. For clarity, we give the details for the S test.
Also, note that we limit our search to the 2-dimensional analysis of the sample $QSO$ since it revealed the most convincing evidence of departure from uniformity with confidence level higher than $99\%$ for the range of $\xi \approx 14.5^\circ \, - \, 17.5^\circ$.

In Section~\ref{sec:uniformity_in_JVAS}, local statistics $S_i$ were computed for each nearest-neighbour group and these statistics have been computed for each simulated dataset. We attribute to each central source $i$ the quantity $s_i$ which tells how much the corresponding group of nearest neighbours contributes to the global statistics $S_D$. This quantity is defined as $s_i = \left( {\left\langle S_i \right\rangle - S_i^\star} \right)/{2\sigma_i}$, where $S_i^\star$ is the statistics obtained for the observed dataset (see Eqs.~\ref{eq:d_theta} and~\ref{eq:S_D}) and where $\left\langle S_i \right\rangle$ and $\sigma_i$ are the mean and the standard deviation of this statistics evaluated over the whole set of simulations. The bigger the value of $s_i$, the more the group contributes to $S_D$.
A group of nearest-neighbour objects is considered as contributing significantly to $S_D$ if $s_i \geq s_c$, for a given threshold $s_c$.

Of course we shall search for the identification of the most significant groups, i.e. consider the $s_i$ quantities computed with the parameter $n_v$ chosen such that $S_D$ is the smallest (see Table~\ref{tab:Summary_SZtest}).
To visualize the sky location of the most significant groups we produce maps which highlight their central sources. Note that these maps do not critically depend on the choice of $n_v$.
These maps are equal area Schmidt projection (e.g. \citealt{Fisher-Lewis-Embleton1987}) of the northern hemisphere (in equatorial coordinates). This choice is suitable for the considered dataset as it covers only positive declinations.
We also plot (the grey bold lines) the limits of the A1 and A3 windows defined in Section~\ref{sec:VisibleWindows}.
Let us insist on the fact that only the northern part of these limits are shown. The A1 and A3 windows defined from the analysis at optical wavelengths extend to the South hemisphere which is not displayed here.

\medskip

\begin{figure}
\begin{center}
\includegraphics[width=\linewidth]{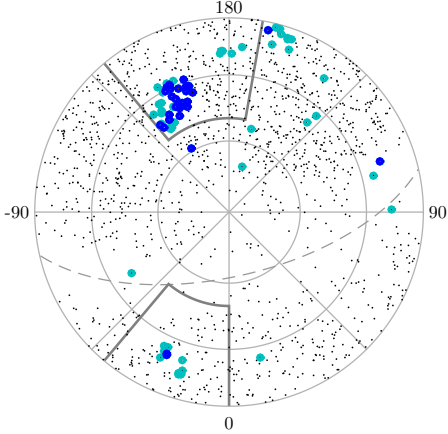}
\caption{Identification map for the sample $QSO$ using the S test in 2D. Parameters are fixed to $n_v=80$ (see Fig.~\ref{fig:2Danalysis-fullSample} (Middle)) and two threshold values: $s_c=2.5$ (lighter points) and $s_c=3.0$ (darker points), respectively cyan and light blue (color on-line). Identification maps are equal area Schmidt projection of equatorial coordinates. Only the equatorial north hemisphere is displayed with the north pole at the centre of the map. Grey circles are parallels of declinations $0^\circ$, $30^\circ$ and $60^\circ$ and grey diagonals are meridians of right ascensions being multiple of $45^\circ$. The curved dashed line is the Galactic equator, the North and the South Galactic caps being respectively above and below the line. Grey bold lines are northern boundaries of the A1 and A3 regions of optical polarization alignments (see text). Small black dots are locations of the $1450$ sources of the sample. Highlighted sources are objects for which corresponding neighbours show a polarization PA distribution that is unlikely due to chance.}
\label{fig:S2D1450_identif}
\end{center}
\end{figure}

The identification map corresponding to the S test (in 2D) for the sample $QSO$ with the parameters $n_v=80$ and $s_c=2.5$ is shown in Fig.~\ref{fig:S2D1450_identif}.
As one can see, the highlighted central sources cluster in three or four groups along with other more sparse and/or isolated locations.
When pushing $s_c$ up to $3$ (darker points), only the cluster with right ascension $\alpha \sim 206^\circ$ and declination $\delta \sim 38^\circ$ remains.
Following this analysis, it is likely that the significant departure from uniformity in this sample is due to polarization alignments in a few groups of QSO's.
It is intriguing that two of them are found in the A1 and A3 windows.

\begin{figure}
\begin{center}
\includegraphics[width=\linewidth]{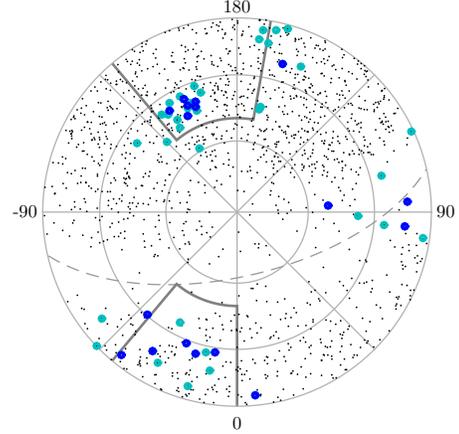}
\caption{Identification map for the sample $QSO$ using the Z test in 2D. Parameters are fixed to $n_v=140$ (see Fig.~\ref{fig:2Danalysis-fullSample} (Bottom)) and $s_c=1.65$ two threshold values: $s_c=1.65$ (lighter points) and $s_c=1.75$ (darker points), respectively cyan and light blue (color on-line).  Please note that the thresholds $s_c$ for the S and Z tests do not refer to the same quantities and have thus different values.}
\label{fig:Z2D1450_identif}
\end{center}
\end{figure}

In order to put the latter identification of aligned regions on stronger grounds, we may use other tests.
The Z test also reveals significant non-uniformity. For the $QSO$ sample and the parameter value $n_v=140$, it leads to the map shown in Fig.~\ref{fig:Z2D1450_identif} which is in relatively good agreement with Fig.~\ref{fig:S2D1450_identif} although it shows more scattered clusters.

Also, we find relevant to proceed to a complementary identification using the density test of \cite{Pelgrims-Cudell2014}.
As this test does not depend explicitly on the number of nearest neighbours (while it is encapsulated within the statistics), local groups can be defined by a physical angular scale, denoted $\Omega$.
In order to carry out an identification as close as possible to those produced with the S and the Z tests, we found necessary to split each sample in its two Galactic hemispheres to determine the physical scale at which local groups have to be defined. Indeed, the density of the data points in the North Galactic cap and the South Galactic cap are different.
In the sample $QSO$, the typical angular separation corresponding to $n_v=80$ is $\xi \approx 17^\circ$ and corresponding to $n_v=140$ is $\xi \approx 24^\circ$ (cf. Fig.~\ref{fig:2Danalysis-fullSample} (Top)).
However, by splitting the sample in its northern and southern Galactic parts, we obtain $\xi_N \approx 16^\circ$ and $\xi_S \approx 23^\circ$ for $n_v=80$ and $\xi_N \approx 21^\circ$ and $\xi_S \approx 33^\circ$ for $n_v=140$, respectively. Given these values, we decided to define local groups in 2 dimensions with angular scales $\Omega_N = 20^\circ$ for the North and $\Omega_S = 30^\circ$ for the South.

\begin{figure}
\begin{center}
\includegraphics[width=\linewidth]{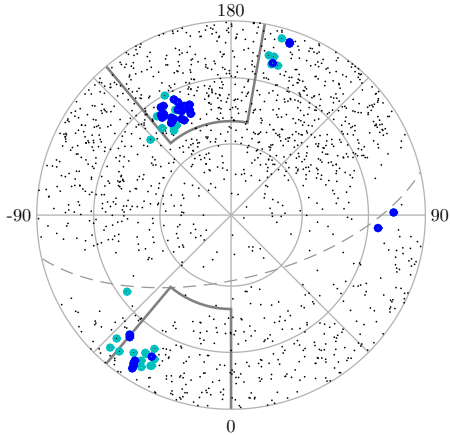}
\caption{Identification map for the sample $QSO$ with the density test. Parameters are fixed as $\Omega=20^\circ$ and $\eta=40^\circ$ for the North Galactic cap and $\Omega=30^\circ$ and $\eta=50^\circ$ for the South Galactic cap. Highlighted sources correspond to groups showing $p_{min} \leq 10^{-3}$ (lighter points) and $p_{min} \leq 3\,10^{-4}$ (darker points), respectively cyan and light blue (color on-line)}.
\label{fig:PC-1450_identif}
\end{center}
\end{figure}

As we search for the characterization of the polarization PA distribution of each group taken as a whole, we shall not investigate values of $\eta$ (the free parameter of the method) below the respective angular separation, i.e. below the imposed angular scales.
We arbitrarily chose $\eta=40^\circ$ and $\eta=50^\circ$ for the North and the South, respectively. The identification map computed with these parameters is shown in Fig.~\ref{fig:PC-1450_identif}.
We checked the robustness of the map with other pairs of values such as $(\Omega_N,\,\Omega_S) = (15^\circ,\,25^\circ)$ and $(25^\circ,\,35^\circ)$. We also checked the stability of our results using other values of $\eta$. Note that we did not search for the optimal value of $\eta$, i.e. the one which would give the lowest probabilities, as the method undergoes edge effects. We rather spanned the range of $20^\circ$ to $60^\circ$ with step of $5^\circ$ and found consistent maps.

Although a close examination shows discrepancies in the precise locations of central sources of neighbouring groups,
comparison of maps presented in Figs.~\ref{fig:S2D1450_identif}{,}~\ref{fig:Z2D1450_identif} and~\ref{fig:PC-1450_identif} reveals relatively good agreement, especially for the cluster at $\left( \alpha,\,\delta\right) \sim \left(206^\circ,\,38^\circ \right)$.

\medskip

In order to define more precisely the limits of regions of polarization alignment, we proceed as follows.
To each central source corresponds a group of nearest objects (defined via the parameter $n_v$ or $\Omega$).
A highlighted central source is said to form a cluster along with (an)other highlighted source(s) if it belongs to the group of nearest objects of the latter.
A central source is discarded from a cluster if it is not in the neighbourhood of a sufficient percentage of central sources forming this cluster (e.g. $\sim 60\%$).
Reproducing this test for all highlighted objects, we end up with identification of independent clusters.
We finally add to the cluster the nearest neighbouring objects of each central sources, paying attention to duplication.
Although this procedure is rudimentary, it is sufficient for our goal.

We thus end up with three regions for each of the three tests.
We decide to define our final regions as being the intersection of the regions from the different tests.
We report the final regions of alignments in Table~\ref{tab:1450_identif-2D} with some characteristics and the result of the application of the Hawley--Peebles test on their polarization PA distributions.

\begin{table}
\caption{Identified regions from the 2-dimensional analysis of the sample $QSO$ with the S, Z and density tests.}
\label{tab:1450_identif-2D}

\medskip

\begin{tabular}{@{}lcccccc}
\hline
& $n$	&	$(\alpha,\,\delta)_{\rmn{CM}}\,({}^\circ)$	&	$\bar{\xi}\,({}^\circ)$	&
$\xi_{max}\,({}^\circ)$	& $P_{\rmn{HP}}\,(\%)$	&	$\bar{\theta}\,({}^\circ)$	\\
\hline
RN1 &	$108$		&	$(163,\,12)$ 	&	$12$ 	& $21$	&	$0.45$	&	$131$	\\
RN2 &	$191$		&	$(206,\,38)$	&	$14$ 	& $25$	&	$1.17$	&	$42$ 	\\
RS1 &	$116$		&	$(340,\,18)$	&	$15$ 	& $25.2$&	$1.45$	&	$57$	\\
\hline
\end{tabular}

\medskip

Regions are intersections of those given by each test (see text). The two first lines are for the regions located in the North Galactic cap and the third is for the region of the South part. They are named RN1, RN2 and RS1, respectively.
$n$ is the number of members belonging to the region, $(\alpha,\,\delta)_{\rmn{CM}}$ refers to the position of the normalized vectorial sum of the source locations of the region, $\bar{\xi}$ and $\xi_{max}$ are the mean and the maximum value of the angular separations of sources to $(\alpha,\,\delta)_{\rmn{CM}}$.
$P_{\rmn{HP}}$ and $\bar{\theta}$ are the results of the Hawley--Peebles test.
\end{table}

\medskip

As a result, we identified three well-defined regions of the sky in which QSO's show coherently oriented polarization vectors.
Two of these regions are located in the North Galactic hemisphere of the sky and one toward the South.
Considering the southern cap, it is worth to remark that more than $85\%$ of the sources of the sub-sample identified here belong to the A3 window defined from the region of optical alignment discovered by Hutsem\'ekers et al. (\citeyear{Hutsemekers1998},~\citeyear{Hutsemekers-Lamy2001},~\citeyear{Hutsemekers-et-al2005}).
To the North, the identified regions are somehow located at the edges of the A1 window of optical alignment, one outside at low declination and the other inside at high declination. We call them RN1 and RN2, respectively. It is again worth mentioning that more than $70\%$ of the sources of the RN2 sub-sample identified here belong to the A1 window.

It is remarkable that our region RN2 coincides with the main aligned cluster resulting from the independent analysis of \citet{Shurtleff2014}. Consistent with our previous results, we report a stronger alignment than he did as we only consider the species of QSO.

\begin{figure}
\begin{center}
\begin{tabular}{@{}c}
\includegraphics[width=0.87\linewidth]{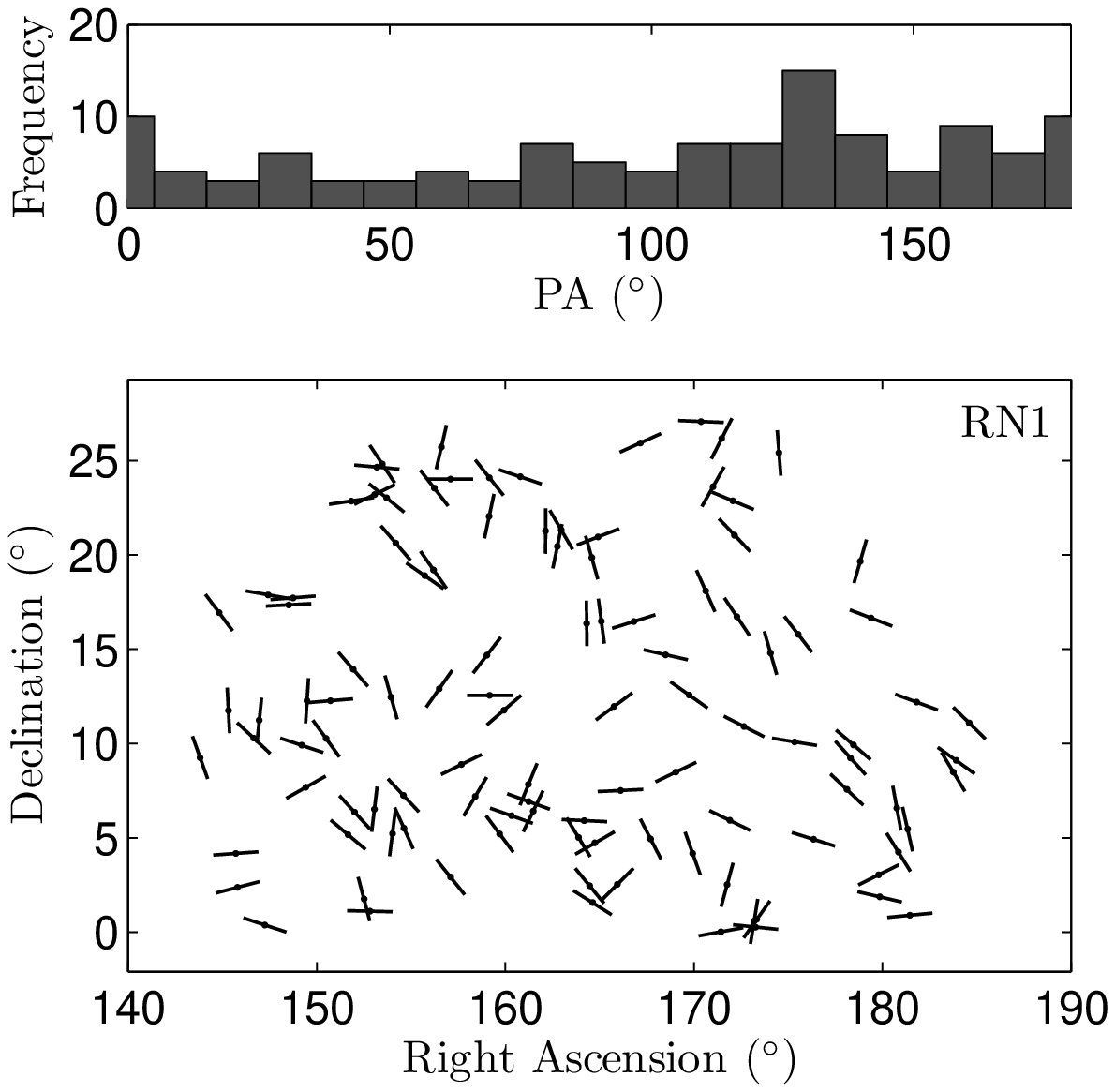}	\\
\includegraphics[width=0.87\linewidth]{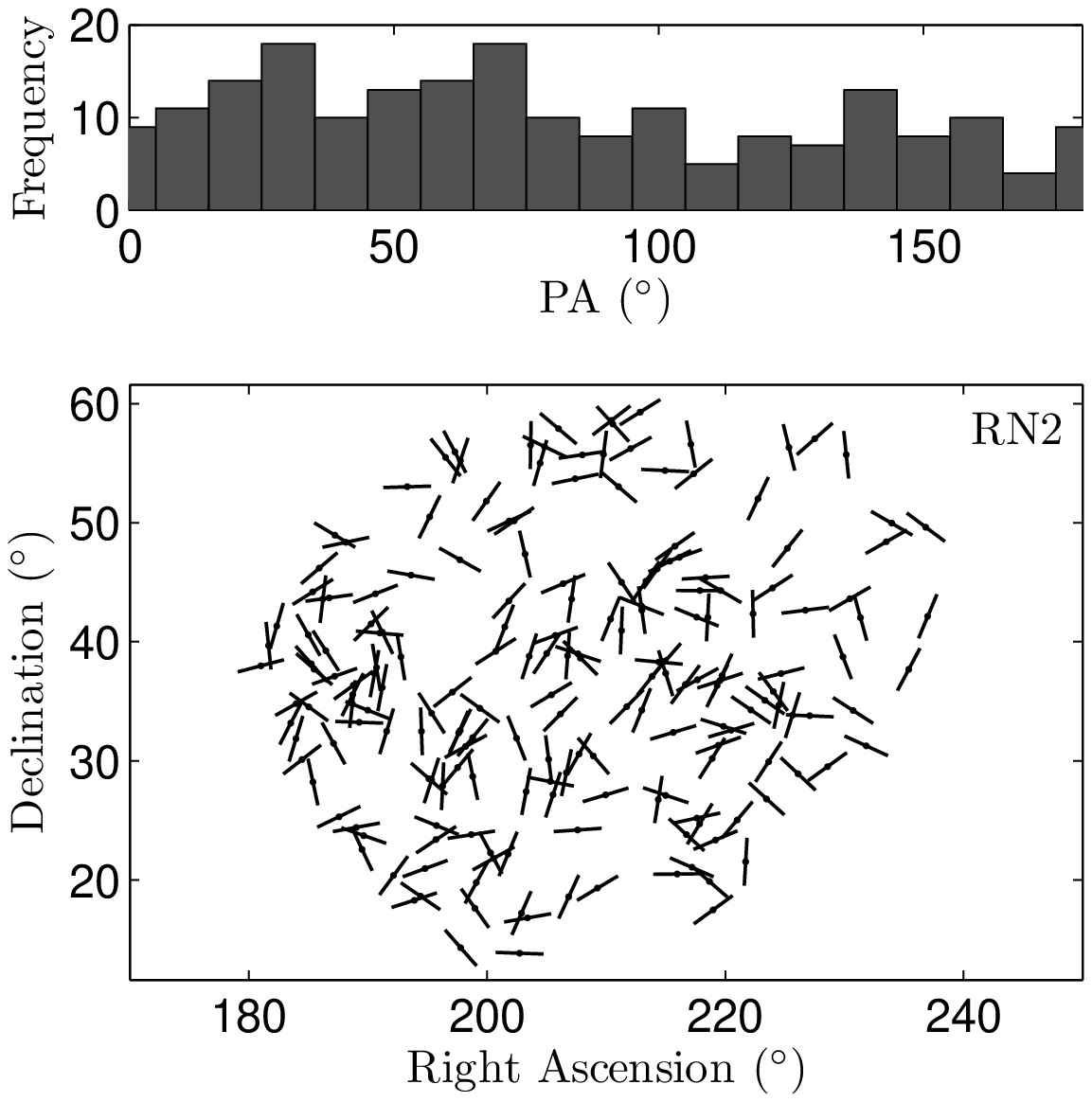}	\\
\includegraphics[width=0.87\linewidth]{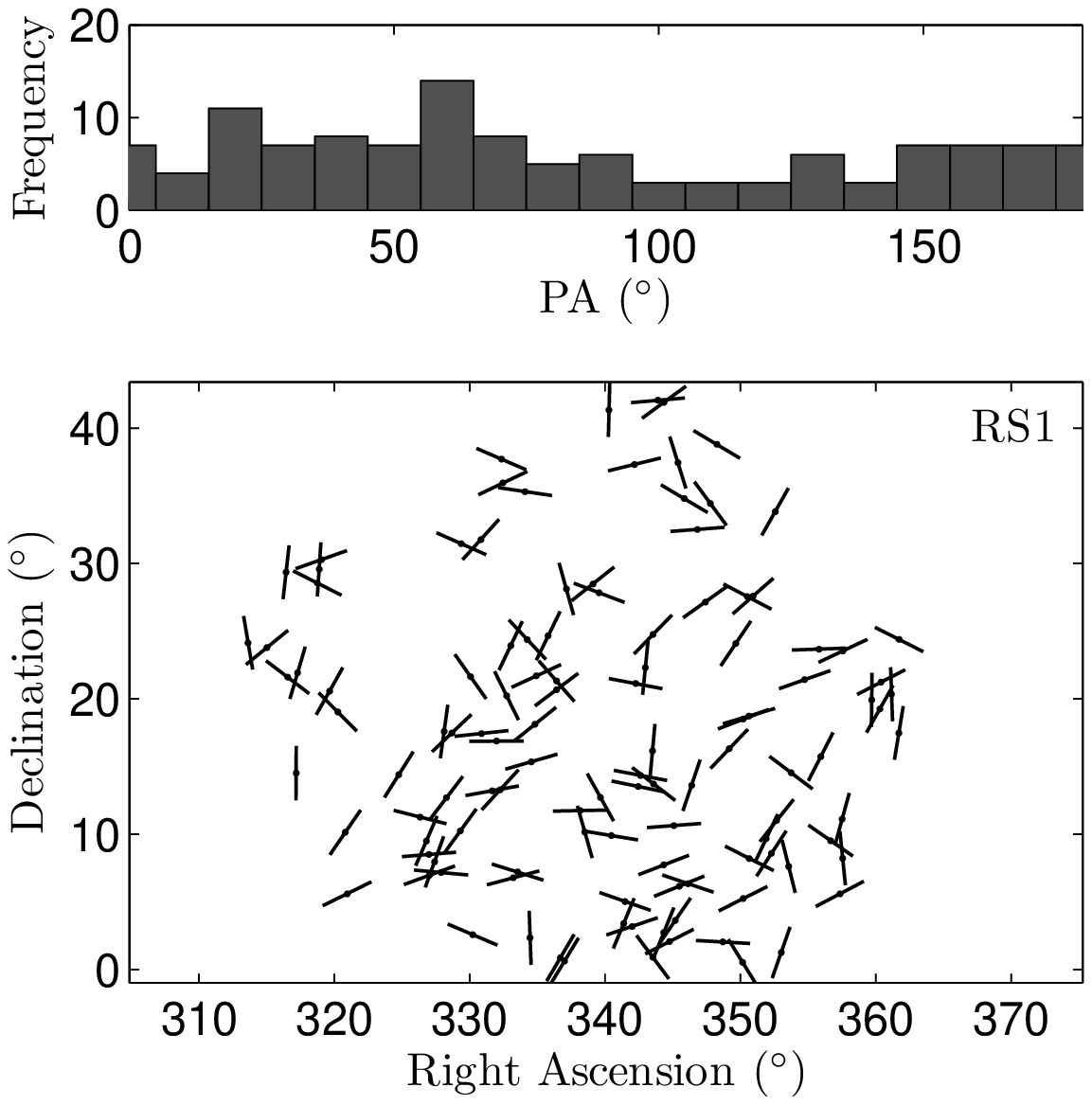}	\\

\end{tabular}
\caption{Maps of polarization vectors in the identified regions along with their corresponding polarization PA histograms. Polarization vectors are normalized to the same length in each map. Top and Middle: the two regions in the North Galactic hemisphere at low and high (equatorial) declination, respectively. Bottom: the region identified in the South Galactic hemisphere. Properties of these regions are given in Table~\ref{tab:1450_identif-2D}.}
\label{fig:IdentifRegionsMap}
\end{center}
\end{figure}

In order to visualize the alignment patterns, we show in Fig.~\ref{fig:IdentifRegionsMap} the equatorial-coordinate maps of the normalized polarization vectors of the identified regions along with their corresponding polarization PA histogram.
Some structures can be spotted out by eye. This is better seen in the region RN1 (see Fig.~\ref{fig:IdentifRegionsMap} (Top)).
The statistical tests used throughout this study do not allow us to better search and characterized such structures. This task is far beyond the scope of this paper and would request dedicated algorithms to compute the likelihood of structures of aligned polarization vectors in a random sample.

\section{Interpreting the results}
\label{sec:DiscussionFinal}
So far we found that the polarization vectors of QSO's being in groups having angular radius of about $20^\circ$ have correlated orientations. We showed that these groups cluster in three independent regions of the sky and that to each of these corresponds a different preferred polarization PA.

\subsection{Are the data contaminated by systematics?}
The preferred angles for the two northern regions are found to have values close to $45^\circ$ and $135^\circ$. These values, being very particular, lead us to consider the possibility that the correlations we found are due to biases in the dataset \citep*[see][]{Battye-Browne-Jackson2008}. This hypothesis, being a priori difficult to reconcile with the local character of the alignment features, could potentially explain that they are better detected with the 2-dimensional analysis than with the 3-dimensional one.
In this sense, and contrary to what \citet{Jackson-et-al2007} and \citet{Joshi-et-al2007} claimed, we also find evidence for a global non-uniformity inside the polarization dataset. Using the Hawley--Peebles test, the probability that the distribution of the $4155$ objects is uniform is found to be $P_{\rmn{HP}}= 2.7\%$ (with $\bar{\theta}\sim 51^\circ$).
This non-uniformity of the overall polarization distribution of the sample $All$ argues for the hypothesis of a biased dataset.
However, consistently with our previous results, this non-uniformity is found to come from the sub-category of QSO as we find $P_{\rmn{HP}}= 1.1\%$ (with $\bar{\theta}\sim 57^\circ$) for this sample and that removing the QSO's from the sample $All$ leads to $P_{\rmn{HP}}= 44.5\%$\footnote{The polarization PA distribution of the sample $RS$ is also in good agreement with uniformity ($P_{\rmn{HP}}= 79.1\%$).}.
This result together with the previous evidence for alignment of QSO's and not for the other species is awkward to reconcile with an observational bias, as there is no reason for a contamination of the polarization data for the species of QSO and not for the others, as we shall see.

\begin{figure}
\begin{center}
\includegraphics[width=\linewidth]{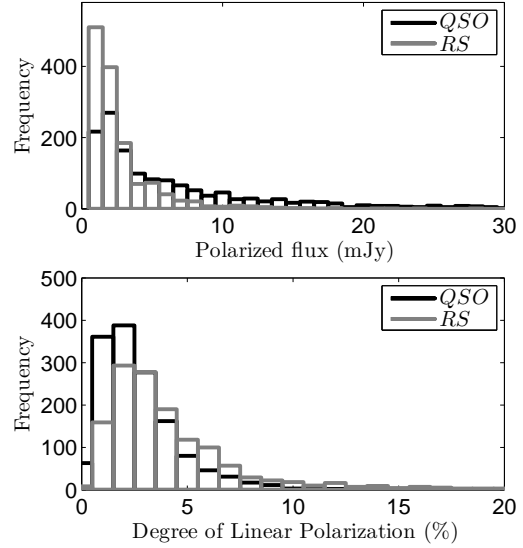}
\caption{Polarized flux (Top) and degree of linear polarization (Bottom) of the samples $QSO$ and $RS$. A two-sample Kolmogorov--Smirnov test reveals that the polarized flux, as well as the degree of linear polarization, of the two samples have a probability much below $1\%$ to be drawn from the same underlying parent distribution.}
\label{fig:QSO-vs-RS-lightPropeties}
\end{center}
\end{figure}

Comparing the properties of the samples $QSO$ and $RS$ (which have a comparable number of objects), we note some differences. As illustrated by the Fig.~\ref{fig:QSO-vs-RS-lightPropeties}, their polarization characteristics at radio wavelengths do not follow the same parent distribution. However, we do not find obvious reasons why the sample $QSO$ would be more affected by observational biases than the sample $RS$ as the QSO sample shows higher total and polarized flux.

The distribution on the sky of the two samples is also different. While the sample $QSO$ is almost homogeneously distributed over the sky, the sample $RS$ is far from being so.
In order to test the possibility that it is the difference of sky distribution that is responsible for the detection of alignments for $QSO$ and not for $RS$, we test the uniformity of the polarization PA of the RS's belonging to the regions of alignments of Table~\ref{tab:1450_identif-2D}.
The overlap is very poor for the regions in the North Galactic hemisphere: only $32$ RS's are found in each of the RN1 and RN2 regions.
To the South, however, there are $165$ RS's in the RS1 region. The Hawley--Peebles test does not reveal departure from uniformity, neither taking RS's alone or mixing them with the QSO's of this region\footnote{This was expected from the analysis of the A3 window (see Section~\ref{sec:PrelResults}).}.
Therefore, while the bad overlap between the $QSO$ and $RS$ samples in the northern regions could explain the difference in the alignment detection, this is not the case to the South. We thus conclude that the difference of sky distribution is unlikely responsible of this difference.

Similarly to an instrumental bias, contamination by foreground polarization would affect more strongly the sample of RS's than the one of QSO's as the polarized flux is globally smaller for $RS$ (see Fig.~\ref{fig:QSO-vs-RS-lightPropeties} (Top)).
This is again in contradiction with what is observed. The contamination by foreground polarization is thus unlikely responsible for the observed correlations of the polarization PA's of QSO's.

\subsection{Are the polarization alignments real?}
As polarization is usually correlated to the morphological axis of the object (e.g., \citealt{Saikia-Salter1988}, \citealt{Lister2001}, \citealt*{Pollack-Taylor-Zavala2003}, \citealt{Smith-et-al2004} and \citealt{Marin2014}), there might be real differences between the classes of QSO and RS. Indeed, the core dominated FSRS's are predominantly quasars or BL Lac objects in which the jet is oriented close to the line-of-sight \citep[see][]{Jackson-et-al2007}. The majority of the sources belonging to the sample $RS$ is thus expected to be BL Lac objects which are thought to be viewed at very small angles to the line-of-sight. Consequently, they are expected to show rapid variations ($<2$ years) of their polarization PA and to be more strongly polarized than quasars, which is what we observe in Fig.~\ref{fig:QSO-vs-RS-lightPropeties} (Bottom). Also, for this class of object, no net correlation between the jet orientation and the polarization PA has been reported \citep[e.g.][and references therein]{Pollack-Taylor-Zavala2003}. These observational facts could bring an explanation to the absence of alignment signatures for the sample of RS's within the hypothesis that polarization alignments reflect morphological-axis alignments of the sources, as supported by the recent discovery at optical wavelengths of such correlation for one of the most largest known quasar group at $z\sim1.3$ \citep{Hutsemekers-et-al2014}.
The latter hypothesis is also reinforced by the discovery of large-scale alignments of the jet position angles of active galactic nuclei in the ELAIS N1 field by \citet{Jagannathan-Taylor2014} and \citet{Jagannathan2014}.
The fact that radio and optical alignments are found in the same parts of the sky also supports a real effect.
In this framework, one would have to compare the alignment patterns observed at optical wavelengths with these at radio wavelengths.
However, given the bad overlap between the sky coverage of the radio and optical catalogues, a detailed comparison is not straightforward.

\section{Conclusion}
\label{sec:Conclusion}
We tested the hypothesis that the polarization position angles are randomly distributed among the FSRS's contained in the JVAS/CLASS 8.4-GHz surveys presented in \citet{Jackson-et-al2007}.
We performed the analysis in two and three dimensions, accounting for the distribution of the sources on the sky (2D) and additionally for their line-of-sight comoving distances (3D).
The polarization orientations of radio quasars ($QSO$ sample) show low probabilities to be consistent with the hypothesis of randomness. This departure from uniformity is likely due to correlations of polarization vectors of QSO's in groups of angular radius of about $20^\circ$. A basic identification procedure showed that these groups cluster in three distinct regions of the sky.
Two of them fall in the A1 and A3 windows of the sky where optical polarization alignments were found (Hutsem\'ekers et al. \citeyear{Hutsemekers1998},~\citeyear{Hutsemekers-Lamy2001},~\citeyear{Hutsemekers-et-al2005}). Among sources in the JVAS/CLASS sample, only the class of QSO exhibits such large-scale correlations.
If real, such alignments at radio wavelengths would support the interpretation of alignments at optical wavelengths by spin-axis alignments \citep{Hutsemekers-et-al2014}. However, our findings prove to be difficult to interpret either as resulting of biases in the dataset or as being the signature of a physical effect.
Indeed, one can find arguments for and against each scenario. Among them, the fact that the alignments are more pronounced in 2D than in 3D and that the mean PA's are multiple of $45^\circ$ in some regions would suggest a biased dataset whereas the detection of alignments for one class of object but not for the others and the clustering of aligned sources in a few regions of the sky consistent with those found at optical wavelengths might be seen as the signature of a physical effect.

In conclusion, we highlighted correlations between the radio-polarization vectors of quasars which could demonstrate the presence of the same kind of alignment effect as seen at optical wavelengths, or alternatively, which could demonstrate that the radio polarization catalogue is affected by observational biases and thus cannot be used to study the polarization orientations of flat-spectrum radio sources.
Therefore, the claim by \citet{Joshi-et-al2007} stating that, at radio wavelengths, there is no alignment signature of polarization vectors on cosmological scales of the type found at optical wavelengths should be seen with caution.

More data are clearly needed to assess either the reality of polarization alignments at radio wavelengths or the presence of residual biases in the JVAS/CLASS 8.4-GHz radio polarization samples.

\section*{Acknowledgements}
We thank J.R. Cudell for bringing our attention to the importance of the accomplishment of this new analysis and for discussions.
D.H. is Senior Research Associate at the F.R.S.--FNRS. This work was supported by the Fonds de la Recherche Scientifique -- FNRS under grant 4.4501.05.
This research has made use of the NASA/IPAC Extragalactic Database (NED) which is operated by the Jet Propulsion Laboratory, California Institute of Technology, under contract with the National Aeronautics and Space Administration. This research has made use of NASA's Astrophysics Data System.

%%%%%%%%

\footnotesize{
\bibliographystyle{mn2e}
\bibliography{mn-jour,myReferences}
}

%%%%%%%%

%\newpage

\appendix
\label{appendix}
\section{Statistical tests}
\label{App_StatTest}
To investigate the polarization PA correlations and to address the question of uniformity of their distributions, we use several statistical tests in this paper.
These are the Fourier method of \citet{Hawley-Peebles1975}, the density test of \citet{Pelgrims-Cudell2014} and the S and Z tests of \citet{Hutsemekers1998}. For the sake of completeness and as they are little-known tests, we hereafter present a brief description of those.

\subsection{The Hawley--Peebles Fourier method}
\label{HP_test}
The Hawley--Peebles test \citep{Hawley-Peebles1975} is one of the common methods used to study the alignments of galaxies. It is based on fitting the observed distribution of PA's by a model of the form $N(\theta_i) = \bar{N} \left(1+\Delta_1 \cos 2\theta_i +\Delta_2 \sin 2\theta_i \right)$ where $\bar{N}$ is the mean of the number of objects per bin and $N(\theta_i)$ is the observed number of objects in the bin centred in $\theta_i$. The number of bins is a free parameter.
$\Delta_1$ and $\Delta_2$ are the coefficients of the wave model which describe the degree of deviation of the distribution from being uniform. If the PA's are not uniformly distributed the mean position angle is given by $\bar{\theta}=({1}/{2})\arctan\left({\Delta_2}/{\Delta_1} \right)$.
A good measure of departure from uniformity is the total amplitude $\Delta^2={\Delta_1}^2+{\Delta_2}^2$. As easily understood, the bigger the value of $\Delta$, the less uniform the distribution.
The probability that the total amplitude exceeds by chance a given value of $\Delta$ is computed to be approximately $P_{\rmn{HP}}=\exp\left( -n\Delta^2 / 4 \right)$ where $n$ is the number of objects in the sample.
However, as far as small samples are considered, random simulations are required as the distribution of $\Delta$ differs from a normal Gaussian and as this approximate relation fails far out in the wings \citep[see][for a detailed discussion]{Godlowski2012}. For the simulated samples, polarization PA's are uniformly generated and distributed among the sources. The probability $P_{\rmn{HP}}$ is then simply given by the percentage of random realizations having a $\Delta$ value higher or equal than that of the data.
In this work, we tested all reported probabilities using random simulations and we only found marginal differences compared to those given by the approximate relation. These are actually smaller than the variations caused by the choice of the number of bin.
We decided to report only the probabilities computed through the Gaussian approximation.
It is worth to mention that this statistical test is also dependent of the coordinate system in which PA's are defined. However, as we use this test to study the uniformity of the polarization PA distributions inside relatively small regions of the sky and that the declinations of these regions are not too high, the changes are expected to be small.

\subsection{The density test of Pelgrims \& Cudell (2014)}
\label{PCdensity_test}
The intrinsically coordinate-invariant statistical test developed in \citet{Pelgrims-Cudell2014} introduces a polarization space (a 2-sphere) and treats polarization as points on it.
The study of the density of these points allows a direct highlight of the direction for which an unexpected over-density is observed. For a given sample of sources a preferred direction is thus automatically allocated, to which corresponds a semi-analytically computed probability. This \textit{local} probability, denoted by $p_{min}$ and being the p-value of a Poisson-Binomial distribution,
gives the likelihood that the observed density toward the given direction is due to chance, i.e. assuming a uniform distribution for the polarization PA's.
A \textit{global} probability, computed through Monte Carlo treatment, is also introduced and is denoted by $p^\sigma$. For a given sample, this global probability answers the question of the likelihood of a detected significant over-density without concern of its direction.
We refer to \citet{Pelgrims-Cudell2014} and \citet{Pelgrims2014} for a complete description of this new test and for a discussion.
Note however that throughout the study of the density of polarization points on the sphere, spherical caps having half-aperture angle $\eta$ are used. This angle is the only parameter of this test and is strongly related to both the sky distribution of the sources and the strength of the alignment of the polarization vectors. It is therefore necessary to explore a large range of value of $\eta$. However, as this is the case with the bin width value for simple binomial tests on histograms, this method undergoes edge effects. A search for the very optimal value of $\eta$ is therefore ambiguous.

During our analysis (e.g. in Section~\ref{sec:VisibleWindows}) we report the mean position angle $\bar{\theta}_{\rmn{PC}}$ for a given distribution. It is computed with respect to the spherical basis vectors $\left(\bmath{{e}_\phi}, \, -\bmath{{e}_\theta} \right)_{\rmn{CM}}$ at the location given by the normalized vectorial sum of the source positions. This quantity has a meaning only when the maximum angular separation between studied sources is not large and when the position angle distribution is not uniform.

\subsection{S and Z statistical tests}
\label{SandZ_tests}
\subsubsection{The S test}
The S test was developed by Hutsem{\'e}kers (1998) in order to detect and statistically characterize the alignment features of polarization vector orientations which were first visually detected.
This test is based on dispersion measures of PA's for groups of $n_v$ neighbouring sources among the sample. For each object, the quantity 
\begin{equation}
d(\theta)=90-\frac{1}{n_v}\sum_{k=1}^{n_v} |90-|\theta_k - \theta||\, ,
\label{eq:d_theta}
\end{equation}
is computed, where the $\theta_k$'s are the polarization PA's of the objects of the group, including the central one, in degree.
For the object $i$, the mean dispersion of the PA's of the $n_v$ objects is computed to be the minimum value of $d(\theta)$ and is denoted $S_i$.
If $n$ is the size of the whole sample under consideration, the statistic with the free parameter $n_v$ is defined as
\begin{equation}
S_D = \frac{1}{n} \sum_{i=1}^{n} S_i \, .
\label{eq:S_D}
\end{equation}
$S_D$ measures the concentration of angles for groups of $n_v$ objects close to each other in space (in 2 or 3 dimensions, as we shall see later). If the polarization vectors are locally aligned, the value of $S_D$ will be smaller than in the case where the PA's are distributed following a uniform distribution on the objects.

The significance level (S.L.) to which one may assign the observed alignment to chance is evaluated through simulated datasets. Those are created by generating PA's according to a uniform distribution or by shuffling the observed PA's among the sources of the sample. This second procedure is found to be more appropriate to the detection of correlations between polarization PA's and the source locations \citep[see,][]{Hutsemekers1998}. This procedure is chosen in this work each time we resort to dataset generation for the S and Z tests.
The percentage of simulations with a value of $S_D$ lower than that of the data gives the S.L., i.e., the probability that the distribution of position angles is due to chance.

\subsubsection{The Z test}
The Z test is a non-parametric test originally introduced by Andrews \& Wasserman \citep{Bietenholz1986} to quantify the correlation between the PA's and the position of sources on the sky. Its basic idea is to compute for each object $i$, the mean direction $\bar{\theta}_i$ of its $n_v$ neighbours, excluding this time the central object $i$, and to compare this local average to the actual polarization PA of the object $i$, namely, $\theta_i$.
Specifically, the PA's of the $n_v$ nearest neighbours around each object $i$ but excluding the latter, are used to computed the mean resultant vector
\begin{equation}
\bmath{Y}_i = \frac{1}{n_v} \left(\sum_{k=1}^{n_v} \cos 2\theta_k ,\, \sum_{k=1}^{n_v} \sin 2\theta_k \right)\, ,
\label{eq:Y_i}
\end{equation}
where the factor $2$ accounts for the axial nature of the polarization.
The mean direction $\bar{\theta}_i$ is given by the normalized mean vector $\bmath{\bar{Y}}_i$ through
\begin{equation}
\bmath{\bar{Y}}_i = \left(\cos 2\bar{\theta}_i, \, \sin 2\bar{\theta}_i \right) \, .
\label{eq:mean_Y_i}
\end{equation}
A measure of the closeness of $\theta_i$ to $\bar{\theta}_j$ is the inner product $D_{i,j}=\bmath{y}_i \cdot \bmath{\bar{Y}}_j$, where $\bmath{y}_i = \left( \cos 2\theta_i,\, \sin 2\theta_i \right) $.
If the PA's are correlated to the source positions, then, on average, $\theta_i$ will be closer to $\bar{\theta}_{j=i}$ than to $\bar{\theta}_{j\neq i}$ which, in turn, implies $D_{i,j}$ to be larger for $i=j$ than for $i \neq j$.
An approximately normally distributed statistic $Z_c$ is computed from these quantities as follow \citep[cf.][]{Bietenholz1986}:
\begin{equation}
Z_c=\frac{1}{n} \sum_{i=1}^{n} Z_i 
\label{eq:Z_c}
\end{equation}
where
\begin{equation}
Z_i=\frac{r_i-\left(n+1\right)/2}{\sqrt{n/12}}
\label{eq:Z_i}
\end{equation}
and where  $r_i$ is the rank of $D_{i,j=i}$, when the $D_{i,j=1,n}$'s are sorted in increasing order and $n$ is the size of the studied sample.
As correlations of PA's with source positions induce large $D_{i,j=i}$, $Z_c$ is expected to be larger than zero if polarization vectors are coherently aligned.

Despite the fact that the $Z_c$ statistic is approximately normally distributed, simulations are needed to evaluate the S.L. because of the mutual dependence of the $D_{i,j=1,n}$,  especially for large values of $n_v$. Therefore, as for the S test, the S.L. is given by the percentage of simulations that show greater value of $Z_c$ than the one corresponding to the observations.

A modification of the Andrews \& Wasserman test was proposed by \citet{Hutsemekers1998}. Instead of considering the inner product with $\bmath{\bar{Y}}_j$, he proposed to compute $D_{i,j}=\bmath{y}_i \cdot \bmath{Y}_j$. This definition gives more weight to the groups of sources having similar PA values.
Indeed, aligned polarization vectors imply a large norm of $\bmath{Y}_i$ which provides a natural measure of the dispersion of the position angles and leads to a large $D_{i,j}$.
Apart from this variation, the modified statistic is computed in the same way as the original test of Andrews \& Wasserman.
In this work, we use this modified version when we refer to the Z test as it is more sensitive to local alignments and thus, more adapted to the search for such features.

\subsubsection{Parallel transport for coordinate-invariant statistics}
The PA's of quantities which are transverse to the line-of-sight to their corresponding sources, i.e. projected in the planes orthogonal to these directions, are dependent of the coordinate system in which the source positions are reported. As the changes of PA values depend on the source positions, results of statistical tests change from coordinate system to coordinate system. In order to overcome this coordinate dependence, \citet{Jain-Narain-Sarala2004} modified the S and Z tests introduced above.

Instead of computing statistics directly from the PA's each evaluated with respect to their own meridian, they introduced correction to the PA's that involves the relative sky positions of the sources. In order to compare in a coordinate-invariant manner the PA's of two sources separated on the celestial sphere, \citet{Jain-Narain-Sarala2004} did parallel transport the polarization vector of one source to the other. The parallel transport is performed along the sphere geodesic passing through the positions of the two sources. The idea is thus to add a correction term to the PA values, this correction being simply the difference between the angles that forms the geodesic with one of the basis vectors at the locations of the two sources.
Therefore, while evaluating Eq.~\ref{eq:d_theta} or Eq.~\ref{eq:Y_i} for each central object $i$, the polarization PA's of the $n_v$ neighbouring sources are transformed as
\begin{equation}
\theta_k ' = \theta_k + \Delta_{k \rightarrow i} \, ,
\label{eq:theta_corrected}
\end{equation}
where each $\Delta_{k \rightarrow i}$ is the correction due to the parallel transport of basis vectors from the position of the source $k$ to the position of the source $i$ \citep[see][for further details]{Jain-Narain-Sarala2004}.

Apart from this modification, the statistical tests are performed as explained above\footnote{Note that \citet{Jain-Narain-Sarala2004} used another definition for the statistics $S_D$ when evaluating Eq.~\ref{eq:d_theta}.} and lead to coordinate-invariant results, as wanted.

\subsubsection{One free parameter statistics and physical interpretation}
\label{nv-param}
In both S and Z tests, the number of nearest neighbours $n_v$ is a free parameter which has to be explored.
Indeed, this parameter is not devoid of physical meaning as it is related to a characteristic scale of the nearest neighbouring groups, in two or three dimensions. This would be strictly true for a sample uniformly scattered across the whole sky. As the observed samples show deviations from homogeneity and, more importantly, as only a part of the entire celestial sphere is covered, this parameter does not show a straight correspondence with a typical size of groups. A dispersion is naturally expected. Nevertheless, if correlations between polarization orientations occur for a typical scale or if some sub-samples well delimited -in space- present such alignments, it is clear that the S.L. will be smaller for the corresponding value of $n_v$ than for others.
Therefore, to test the uniformity of the polarization orientations and explore their characteristics, it is necessary to estimate the S.L. across a wide range of values of $n_v$.
To build the groups of nearest neighbouring objects, the relative distances between sources is required.
For this purpose, and assuming a flat and isotropic Universe, we consider the line-of-sight comoving distance $r$ to the observer \citep[e.g.][]{Hogg2000}.
%\begin{equation}
%r(z) = \frac{c}{H_0} \int_0^z \frac{dz'}{\sqrt{1 + \Omega_M \left( (1+z')^3 - 1 \right)}}
%\label{eq:D_Comoving}
%\end{equation}
%where $c$, $H_0$ and $\Omega_M$ are respectively the speed of light, the Hubble parameter and the matter density parameter.
In agreement with \citet{Ade-et-al2014}, our calculation are made with $H_0 = 68 \, \rmn{ km \, s}^{-1} \rmn{Mpc}^{-1}$ and $\Omega_M = 0.31$.

%From Eq.~\ref{eq:D_Comoving}, the rectangular coordinates of each source are evaluated in a flat Universe trough
%\begin{equation}\label{eq:xyz-coordinates}
%	\begin{gathered}
%		{x =r \cos \delta \cos \alpha}  \\
%		{y =r \cos \delta \sin \alpha}  \\
%		{z =r \sin \delta}
%	\end{gathered}
%\end{equation}
%where $\delta$ and $\alpha$ denote here also the declination and the right ascension of the object in the equatorial coordinate system.
The relative distances between sources are then simply computed through the use of their rectangular coordinates and the groups of nearest neighbours constructed.
A three dimensional analysis is obviously only feasible for samples for which redshift measurements are available. Nevertheless, a two dimensional one, considering nearest neighbours on the celestial sphere rather than in the 3D space, is not devoid of interest and is applicable for all samples imposing $r=1$.

\label{lastpage}

\end{document}